\documentclass[reprint, citeautoscript, superscriptaddress, prb,longbibliography, amssymb, amsmath]{revtex4-2}

\usepackage{graphicx}
\usepackage{dsfont}

\usepackage[colorlinks=true,linkcolor=red,urlcolor=magenta,citecolor=blue,pdfusetitle]{hyperref}

\usepackage{braket}

\UseRawInputEncoding

\begin{document}

\title{Quantum master equation from the eigenstate thermalization hypothesis}
\author{Peter O'Donovan}
\email[]{peodonov@tcd.ie}
\affiliation{School of Physics, Trinity College Dublin, College Green, Dublin 2, D02 K8N4, Ireland}

\author{Philipp Strasberg} 
\affiliation{F\'isica Te\`orica: Informaci\'o i Fen\`omens Qu\`antics, Departament de F\'isica, Universitat Aut\`onoma de Barcelona, 08193 Bellaterra (Barcelona), Spain}
\affiliation{Instituto de F\'isica de Cantabria (IFCA), Universidad de Cantabria--CSIC, 39005 Santander, Spain}

\author{Kavan Modi} 
\affiliation{School of Physics \& Astronomy, Monash University, Victoria 3800, Australia}

\author{John Goold}
\affiliation{School of Physics, Trinity College Dublin, College Green, Dublin 2, D02 K8N4, Ireland}
\affiliation{Trinity Quantum Alliance, Unit 16, Trinity Technology and Enterprise Centre, Pearse Street, Dublin 2, D02 YN67, Ireland}

\author{Mark T. Mitchison} 
\email{mark.mitchison@tcd.ie}
\affiliation{School of Physics, Trinity College Dublin, College Green, Dublin 2, D02 K8N4, Ireland}
\affiliation{Department of Physics, King’s College London, Strand, London, WC2R 2LS, United Kingdom}

\date{\today}

\begin{abstract}
    We use the eigenstate thermalization hypothesis to derive a quantum master equation for a system weakly coupled to a chaotic finite-sized bath prepared in a pure state. We show that the emergence of Markovianity is controlled by the spectral function of the ETH and that local detailed balance emerges in the Markovian regime for a broad class of pure bath states. We numerically verify this result by comparing the master equation to dynamics computed using exact diagonalization of a chaotic Hamiltonian. We also compare the master equation to exact dynamics for an integrable bath and find that at finite size they strongly disagree. Our work puts forward eigenstate thermalization as a foundation for open quantum systems theory, thus extending it beyond ensemble bath preparations to chaotic many-body environments in generic pure states.    
\end{abstract}
\maketitle

\section{Introduction}

Master equations are important for studying thermalization and irreversibility in open quantum systems (OQS). In OQS theory, these phenomena are explained by the interaction between a system and some inaccessible environment. The Redfield and Lindblad master equations describe the dynamics of systems which are weakly coupled to an environment and are frequently discussed in textbooks on OQS theory \cite{rivas_open_2012_real, breuer_theory_2007_real, PRXQuantum.5.020202}. The derivation of these equations require a variety of assumptions to be made about the Hamiltonian and state of the bath. In this work, we focus on the widely used assumption that the bath is prepared in a Gibbs ensemble.

This assumption is not always justified because a bath prepared in an arbitrary nonequilibrium state will not thermalize in general. This is particularly clear for pure bath states, which never evolve to a Gibbs ensemble under unitary dynamics. Additionally, local thermalization will not be guaranteed for many of the frequently used models considered in OQS theory, such as free fermions, free bosons and simple spin models. Instead these non-chaotic models are expected to locally evolve towards a generalized Gibbs ensemble \cite{vidmar_generalized_2016_real}. Even if such non-chaotic systems are somehow prepared in a standard Gibbs state, the ensuing dynamics may not satisfy the requirements for a weak-coupling master equation to be valid~\cite{TASAKI2007631,davies_markovian_1974, parra-murillo_open_2021}. Therefore, 
what justifies the use of a Gibbs ensemble for the bath has remained an unanswered question within OQS theory.

This issue becomes relevant when considering isolated quantum systems, such as ultracold atomic gases, that can be described by a pure state. It is now possible to embed impurity probes in these gases \cite{PhysRevLett.109.235301, doi:10.1126/science.1240516, PhysRevX.10.011018, PhysRevLett.129.120404}, as well as control their interaction with the gas \cite{doi:10.1126/science.aaf5134, Baroni_2023}. It has also been possible to study impurities coupled to strongly interacting environments \cite{PhysRevLett.103.223203, Kohstall_2012}. These impurities are most naturally described using OQS theory with a pure bath preparation. This motivates a master equation derivation that does not assume a thermal ensemble for the environment, complementing other recent efforts to extend OQS theory beyond standard assumptions, e.g.~to finite sized \cite{Gemmer_Bartsch_2008, esposito_quantum_2003, riera-campeny_quantum_2021_real, riera-campeny_open_2022, moreira_stochastic_2023}, non-Markovian \cite{de_vega_dynamics_2017} and non-thermal \cite{Gemmer_Breuer_2006, pereyra_random-matrix_1991} baths.

To tackle this problem, we use methods developed to explain thermalization in isolated quantum systems: specifically, quantum chaos theory. Chaotic many-body Hamiltonians are known to obey the eigenstate thermalization hypothesis (ETH), which states that energy eigenstates yield thermal expectation values for local operators (see Sec.~\ref{sec:set_up}). It has been proven that the ETH is a sufficient condition for thermalization in isolated quantum systems prepared in high energy, pure states with subextensive energy fluctuations \cite{deutsch_quantum_1991, srednicki_chaos_1994, srednicki_thermal_1996, srednicki_approach_1999, dalessio_quantum_2016}. Numerical experiments have verified the ETH in chaotic many-body Hamiltonians \cite{rigol_thermalization_2008, dalessio_quantum_2016, deutsch_eigenstate_2018, chan_eigenstate_2019, PhysRevE.102.062113} and in Floquet dynamics \cite{garratt_local_2021, chan_eigenstate_2019, fritzsch_eigenstate_2021, pappalardi2024eigenstatethermalizationfreecumulants, kim_testing_2014}. The ETH and random matrix theory have been used to understand nonequilibrium phenomena such as transport properties \cite{alhassid_statistical_2000, beenakker_random-matrix_1997, wang_eigenstate_2022, brenes_eigenstate_2020, roy_anomalous_2018, PhysRevA.107.022220, PhysRevA.105.L040203, zhang2024emergencesteadyquantumtransport} and have been used to derive fluctuation-dissipation theorems \cite{iyoda_fluctuation_2017, jin_eigenstate_2016, nation_quantum_2019_real,noh_numerical_2020, schonle_eigenstate_2021,  mitchison_taking_2022}. The ETH is expected to hold for local operators or sums of local operators \cite{PhysRevLett.122.070601, PhysRevX.8.021026} and so it should be applicable to models with local, short-ranged interactions, making it a potentially useful resource in understanding quantum impurity problems.

Properties of many-body systems, such as information scrambling and the Lieb-Robinson bound, have been used to assist in deriving master equations \cite{shiraishi_quantum_2024_real}. The ETH in particular has been used to derive classical master equations \cite{dalessio_quantum_2016, Strasberg_master_eq_pure} and equations describing the dynamics of local operators in chaotic models \cite{nation_quantum_2019_real, Strasberg_master_eq_pure}. Additionally, the ETH has been applied to OQS theory, but generally only ensemble initial bath preparations are considered \cite{fialko_quantum_2015, parra-murillo_open_2021, wang_thermalization_2022_real, wang_internal_2017_real, wang_interplay_2023_real}. Despite these works, a full derivation of the quantum master equation for pure bath states has not been performed.

In Sec.~\ref{sec:derivation}, we fill this gap by deriving a weak-coupling quantum master equation for a system interacting with a bath that obeys the ETH and is prepared in a pure state. This establishes quantum chaos as a primary condition for the applicability of OQS theory when considering isolated environments. We show that the ETH implies a local detailed balance condition, which allows us to establish a consistent nonequilibrium thermodynamics and contributes to a line of research aiming to understand the implications of chaos and the ETH far from equilibrium \cite{iyoda_fluctuation_2017,iyoda_eigenstate_2022,jin_eigenstate_2016,camalet_joule_2008,steinigeweg_pushing_2014,kim_testing_2014,noh_heating_2019,nation_quantum_2019_real,deutsch_quantum_1991,reimann_typicality_2007,reimann_foundation_2008, reimann_refining_2021,dabelow_typical_2021_real,dabelow_thermalization_2022}. Our master equation gives a quantitative description of thermalization timescales in the weak-coupling regime.
Additionally, generic Hamiltonians are expected to obey the ETH \cite{dalessio_quantum_2016}, so we expect the master equation to hold for most bath dynamics and even for multiple baths.

We numerically verify this master equation in Sec.~\ref{Numerical Results} by comparing it to dynamics computed using exact diagonalization. This is done for a variety of pure bath states such as single eigenstates, random states and product states. We also consider an integrable model which shows disagreement with the master equation for finite sized baths. These results indicate that in finite sized systems, chaos ensures a consistent nonequilibrium thermodynamics for a much broader class of initial bath states than is commonly used in OQS theory.

Units with $\hbar = 1 = k_B$ are used throughout this paper.

\section{Setup}
\label{sec:set_up}

\begin{figure}[t]
    \centering
\includegraphics[width=0.5\textwidth]{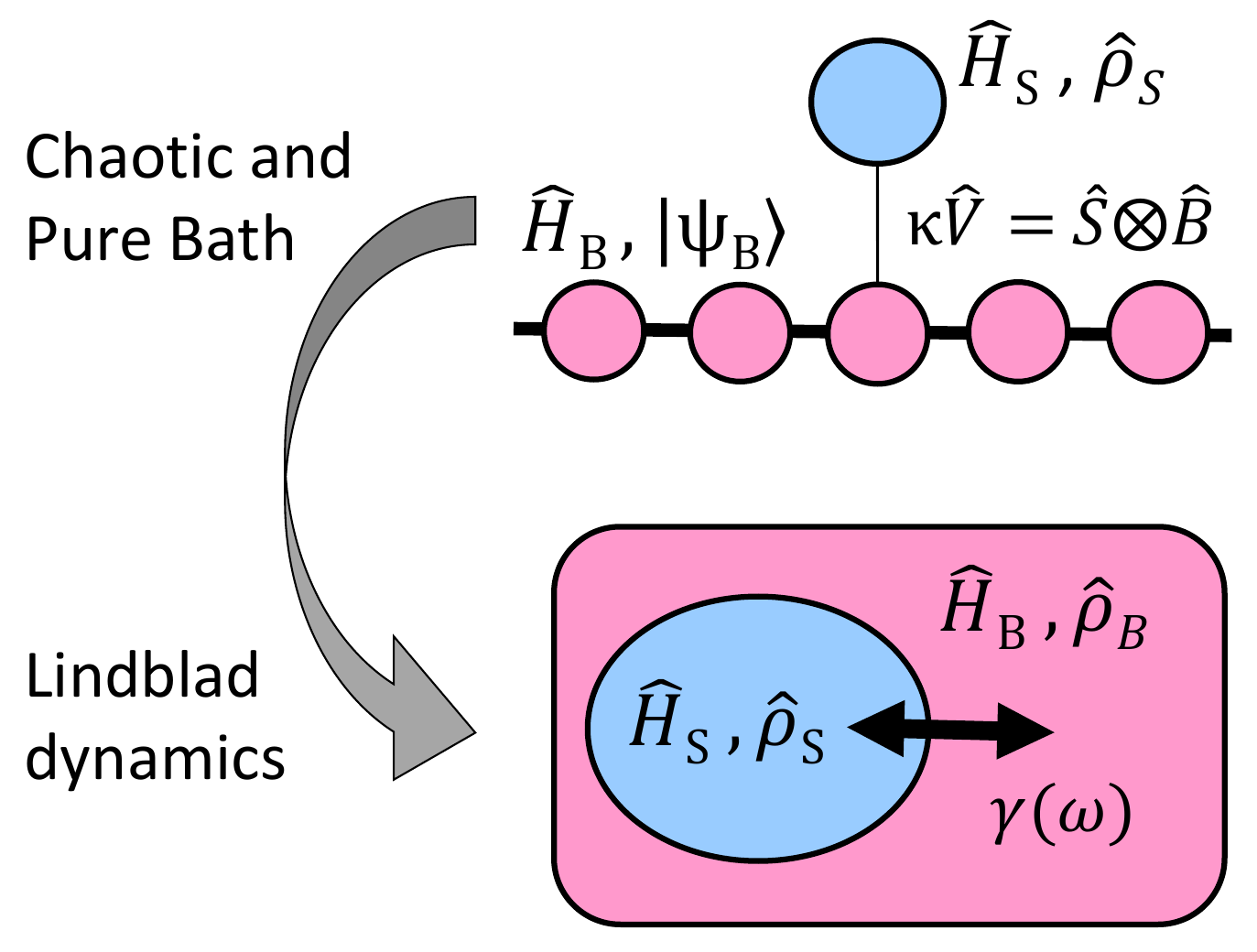}
    \caption{Diagram of physical set-up with a system weakly coupled to the bath which is prepared in a pure state. Due to the locality of this coupling and that the bath obeys the ETH, we can show that the system dynamics simplifies to a Lindblad master equation, as if it was coupled to a bath prepared in a thermal state.}
    \label{fig:physical_system}
\end{figure}

OQS theory is concerned with describing the dynamics of a relatively small system of interest that is influenced by its coupling to a large, many-body environment or bath, as sketched in Fig.~\ref{fig:physical_system}. This situation is described by a Hamiltonian of the form
\begin{equation}
\label{Hamiltonian}
    \hat{H} = \hat{H}_{S} + \hat{H}_{B} + \kappa \hat{V}.
\end{equation}
Here, $\hat{H}_{S}$ and $\hat{H}_{B}$ are the Hamiltonians for the system and bath respectively, while $\hat{V}$ describes their mutual coupling with interaction strength $\kappa$. In general, the interaction Hamiltonian can be written as 
\begin{equation}
\label{V_general}
    \hat{V} = \sum_{\mu} \hat{S}^{\mu}\otimes \hat{B}^{\mu},
\end{equation}
with Hermitian system and bath operators $\hat{S}^{\mu}$ and $\hat{B}^{\mu}$. We assume that the OQS comprises some localised degrees of freedom embedded in a spatially extended bath, such as an impurity in a solid-state lattice~\cite{kretinin_universal_2012, pustilnik_kondo_2004} or an ultracold atomic gas~\cite{Astrakharchik2004,Recati2005a,Recati2005b}. The bath operators $\hat{B}_\mu$ are therefore local observables of the many-body bath. We assume throughout that the system-bath coupling $\kappa$ is small enough to be treated as a weak perturbation. 

In standard treatments of OQS theory, the bath is usually taken to be a non-interacting system prepared in a canonical or grand-canonical thermal state~\cite{breuer_theory_2007_real, rivas_open_2012_real, PRXQuantum.5.020202}. Features of the initial system state are thus eventually washed out by thermal uncertainty in the initial bath preparation. Here, in contrast, we consider a bath with strong internal interactions prepared in a pure quantum state $\ket{\psi_B}$, such that the initial density matrix of the model is of the form
\begin{equation}
    \label{product_state}
    \hat{\rho}_{SB}(0) = \hat{\rho}_S(0) \otimes \ket{\psi_B}\bra{\psi_B},
\end{equation}
with $\rho_S(0)$ the (arbitrary) initial state of $S$. For a completely arbitrary pure state $\ket{\psi_B}$, one does not expect the emergence of a consistent nonequilibrium thermodynamics, even at weak-coupling $\kappa$. The dynamics of the system is also not generically expected to be Markovian. In the following, however, we will show that both of these properties occur for a wide class of initial pure states $\ket{\psi_B}$, so long as the bath obeys the eigenstate thermalization hypothesis (ETH). Before proceeding to derive the master equation, however, we first review salient features of the ETH~\cite{ srednicki_chaos_1994, deutsch_eigenstate_2018, dalessio_quantum_2016}; readers familiar with the ETH may wish to skip directly to Sec.~\ref{sec:derivation}.

 The ETH is encapsulated by an ansatz for the matrix elements of a local operator written in the basis of the energy eigenstates. For a local observable $\hat{B}$---having a support that is smaller than its complement~\cite{PhysRevE.97.012140}---and a chaotic Hamiltonian $\hat{H}_B$ whose eigenvalues and eigenstates are $E_n$ and $|n\rangle$ respectively, the ETH ansatz is given by the following ~\cite{srednicki_approach_1999}
\begin{equation}
\label{eq:ETH_ansatz}
    \langle n|\hat{B}|m\rangle = B(E_{nm})\delta_{nm} + e^{\frac{-S(E_{nm})}{2}}f(E_{nm}, \omega_{nm})R_{nm}.
\end{equation}
Above, $R$ is a pseudorandom matrix with elements having zero mean $\overline{R_{nm}} = 0$ and unit variance $\overline{R_{nm}^2} = 1$. This pseudorandomness arises from due to the complexity of eigenstates of the chaotic Hamiltonian. The functions $B(E)$, $S(E)$ and $f(E,\omega)$ are smooth in their arguments with $E_{nm} = (E_n + E_m)/2$ and $\omega_{nm} = (E_m - E_n)$ for fixed system size $L$. The smooth function $B(E)$ equals the expectation value of the local observable $\hat{B}$ in the microcanonical ensemble at energy $E$. Meanwhile, $S(E)$ is the thermodynamic (Boltzmann) entropy and $f(E, \omega)$ is known as the spectral function of $B$. Using the fact that $B$ is Hermitian, the spectral function will obey the condition $f(E, \omega) = f^*(E, -\omega)$ ~\cite{dalessio_quantum_2016}. We consider Hamiltonians $H_B$ that have time-reversal symmetry, so the spectral function is real.

\begin{figure}[t]
    \centering
    \includegraphics[width=0.5\textwidth]{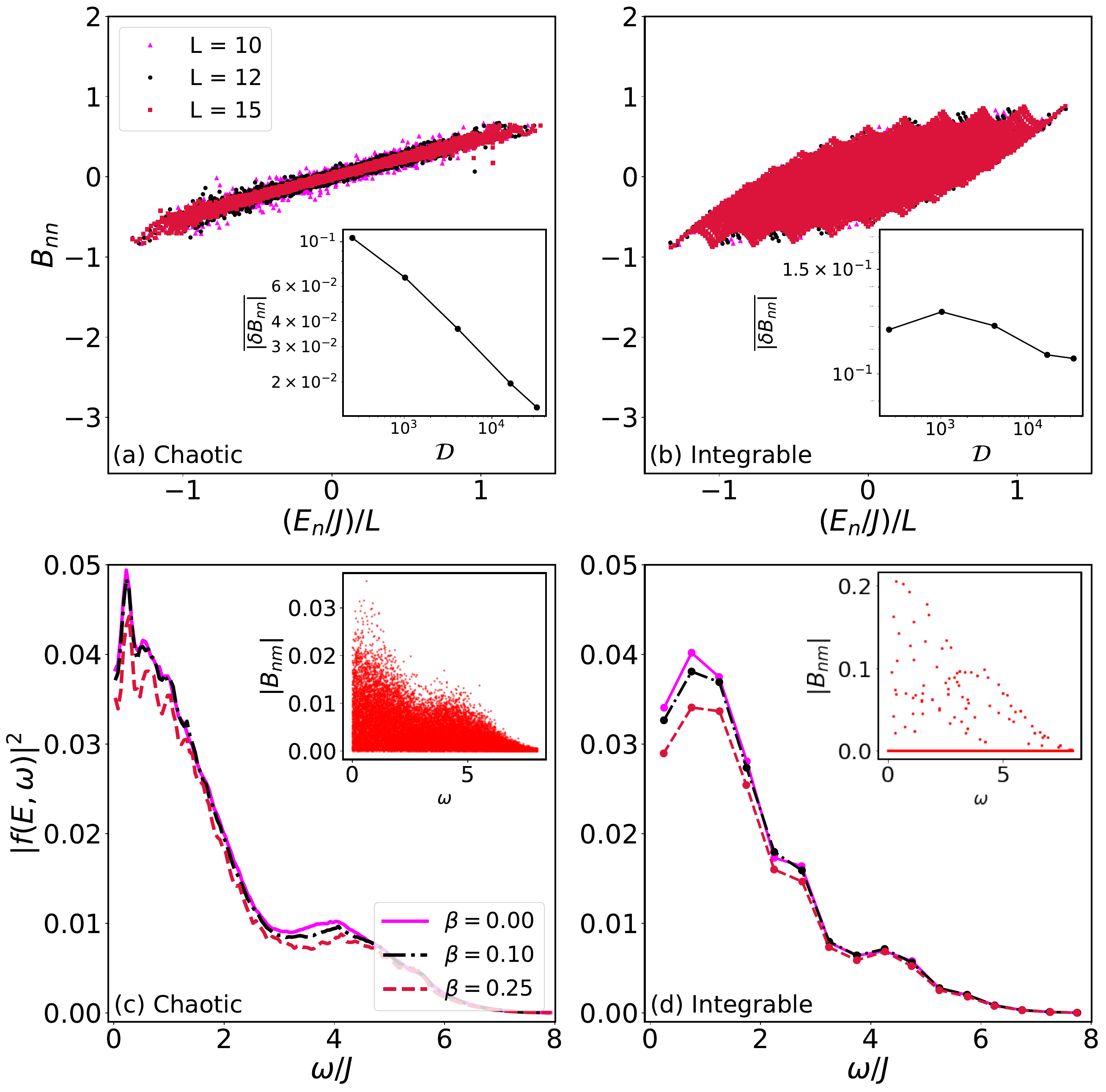}
    \caption{[(a), (b)] Diagonal matrix elements of a local operator ($\hat{B}=\hat{\sigma}_1^x$) in the basis of the energy eigenstates of a chaotic and integrable spin chain, respectively. See Eq.~\eqref{TL_Ising} for details of the Hamiltonian and parameters chosen in each case. Diagonal matrix elements in the chaotic model concentrate around a smooth function of energy as system size $L$ increases. Inset: the eigenstate fluctuations of the diagonals $\overline{|\delta B_{nn}|} = \overline{|B_{n+1 n+1} - B_{nn}|}$ with the average computed over the central 10\% of energies, plotted as a function of the Hilbert space dimension $\mathcal{D}$. The fluctuations in the chaotic case are observed to decay exponentially in system size, while the fluctuations in the integrable remains approximately constant. [(c), (d)] Off-diagonal matrix elements computed for the chaotic and integrable model, respectively. We extract a smooth function of frequency for the chaotic case. Inset: the off-diagonal elements are dense in the chaotic case and have fluctuations as the pseudorandom matrix term in the ETH predicts. The integrable case is mostly sparse which means it cannot be ``meaningfully" described by a smooth function like in the chaotic case.}
    \label{fig:ETH_data}
\end{figure}

The meaning of Eq.~\eqref{eq:ETH_ansatz} is best illustrated with a numerical example. Fig.~\ref{fig:ETH_data} shows data obtained from exact diagonalisation of the Hamiltonian of a chaotic quantum spin chain of length $L$ (see Sec.~\ref{Numerical Results:1} for details). For increasing system size, the diagonal matrix elements of the local operator $B$ concentrate around a smooth function of energy [Fig.~\ref{fig:ETH_data}(a)]. As $L\to\infty$, all energy eigenstates near energy $E$ thus yield the same expectation value, $B(E)$, which must therefore be equal to the microcanonical average over a small energy window. 

Beyond energy eigenstates, the ETH implies that a much larger set of nonequilibrium pure states thermalize to the prediction of the microcanonical ensemble under unitary time evolution~\cite{deutsch_quantum_1991,srednicki_thermal_1996,srednicki_approach_1999, rigol_thermalization_2008}. More precisely, thermalization for bath observables means that
\begin{equation}
    \label{thermalization_time_average}
    \lim_{t\to\infty}\frac{1}{t} \int_0^t dt'\, \langle \psi_B|\tilde{B}(t')|\psi_B\rangle = B(E) + \mathcal{O}(L^{-1}),
\end{equation}
where $\tilde{B}(t) = e^{i\hat{H}_Bt}\hat{B} e^{-i\hat{H}_Bt}$ and $E = \langle \psi_B|\hat{H}_B|\psi_B\rangle$ is the mean energy of the pure state $|\psi_B\rangle$ and $\mathcal{O}$ refers to big O notation. Additionally, temporal fluctuations will be small, so that the expectation value looks thermal at most times, when the bath state has subextensive energy fluctuations $\Delta E/E \propto \mathcal{O}(L^{-\frac{1}{2}})$, with energy variance given by $\Delta E^2 = \langle \psi_B|\hat{H}_B^2|\psi_B\rangle - \langle \psi_B|\hat{H}_B|\psi_B\rangle^2$ \cite{dalessio_quantum_2016}. It is known that a wide variety of pure states satisfy these conditions, including product states and states with exponentially decaying spatial correlations \cite{anshu_concentration_2016_real}.

The second term in Eq.~\eqref{eq:ETH_ansatz} describes the off-diagonal matrix elements of a local operator $\hat{B}$ in the energy eigenbasis. As shown in Fig.~\ref{fig:ETH_data}(c), these elements are generically non-zero but exponentially small in the system size, and are erratically distributed with a variance governed by the smooth spectral function $f(E,\omega)$. This function is related to the Fourier transform of the two-point autocorrelation function of $\hat{B}$, and its smooth properties allow fluctuation-dissipation relations to be derived for pure states~\cite{srednicki_approach_1999, iyoda_fluctuation_2017, jin_eigenstate_2016, nation_quantum_2019_real,noh_numerical_2020, schonle_eigenstate_2021,  mitchison_taking_2022}. The spectral function depends on the operator and the microscopic details of the model; however, it generically displays exponential decay in frequency $\omega$ above some frequency scale $\Delta$~\cite{khatami_fluctuation-dissipation_2013, dalessio_quantum_2016}. This feature is crucial for our derivation of the master equation that follows.

Strikingly different behaviour is observed in integrable systems, which do not obey the ETH. This class includes non-interacting systems, which form the canonical environment models for OQS theory, as well as fine-tuned interacting models with symmetries that render them solvable by Bethe ansatz~\cite{Sutherland_models}. For integrable systems, diagonal matrix elements of local observables in the energy eigenbasis exhibit a significant variance even for large system size [Fig.~\ref{fig:ETH_data}(b)]. This behaviour is expected in integrable systems \cite{santos_localization_2010, PhysRevE.87.012125} and indicates that the long-time evolution of observables does not depend only on mean energy, as in Eq.~\eqref{thermalization_time_average}, but also on other conserved quantities as in a generalized Gibbs ensemble~\cite{vidmar_generalized_2016_real}. 

In non-interacting integrable models, an extensive fraction of the off-diagonal matrix elements vanish due to symmetry constraints~\cite{PhysRevLett.111.050403, zhang_statistical_2022} [Fig.~\ref{fig:ETH_data}(d)]. As a result, it is believed that these off-diagonals cannot be characterised by meaningful smooth spectral functions. For interacting integrable models, however, it has been shown that the off-diagonal matrix elements can be described by smooth functions of frequency in the infinite-temperature regime~\cite{leblond_entanglement_2019, PhysRevE.102.062113}. Nevertheless, the low-frequency behaviour of these functions is starkly different in the chaotic and integrable case~\cite{Brenes2020,Pandey2020}, as it encodes hydrodynamic behaviour in the thermodynamic limit~\cite{mitchison_taking_2022,capizzi_hydrodynamics_2024_real}. Here, in contrast, we focus primarily on the distinction between non-interacting integrable systems --- the typical bath models of OQS theory --- and generic chaotic models. This distinction leads to significant differences in the dynamics of the OQS, as we show in Sec.~\ref{Numerical Results}.

\section{Master Equation Derivation}
\label{sec:derivation}

\subsection{Energy eigenstate bath preparation}
\label{sec:energy_eigenstate}

To derive a quantum master equation from the ETH, we first make the simplifying assumption that the initial bath state is a pure energy eigenstate, $\ket{\psi_B} = \ket{n}$. This assumption is relaxed to consider more general bath preparations in Sec.~\ref{sec:general_state}. To keep the notation simple, we also assume that the interaction Hamiltonian $\hat{V} = \hat{S}\otimes \hat{B}$ has a single term in this section. This assumption already captures many physically relevant scenarios, but more general interactions of the form~\eqref{V_general} are also considered in Sec.~\ref{sec:general_int}. For completeness, we merely summarise the approximations involved in deriving a weak-coupling master equation at second order in coupling $\kappa$; see Ref.~\cite{breuer_theory_2007_real}.

Stationarity of the bath state $|\psi_{B}\rangle = e^{-i\hat{H}_{B}t}|\psi_{B}\rangle$, ensures that its one- and two-point correlation functions are translationally invariant in time:
\begin{equation}
\label{stationarity_cond}
\begin{aligned}
    &\langle \tilde{B}(t) \rangle_0 = \langle \tilde{B}(0) \rangle_0, \; \;\langle \tilde{B}(t)\tilde{B}(t')\rangle_0  = \langle \tilde{B}(t- t')\tilde{B}(0)\rangle_0.
\end{aligned}
\end{equation}
Here, $\tilde{B}(t) = e^{i\hat{H}_Bt} \hat{B} e^{-i\hat{H}_Bt}$ denotes the interaction-picture evolution and $\braket{\bullet}_0 = \braket{\psi_B|\bullet |\psi_B}$ denotes an average with respect to the initial bath state. A pure energy eigenstate is a stationary state with respect to the bath dynamics so this condition will be satisfied.

Using stationarity and the weak-coupling approximation, one may derive a time-convolutionless master equation at second order in the system-bath coupling (e.g.~see Ref.~\cite{breuer_theory_2007_real}, Sec.~9.2), given by
\begin{align}
\label{eq:weak_coupling}
    \frac{\partial \tilde{\rho}_{S}}{\partial t} & = \kappa^2\int^{t}_{0} d\tau\, C(\tau) \left[ \tilde{S}(t - \tau) \tilde{\rho}_{S}(t - \tau)\tilde{S}(t) \right. \notag \\
    & \qquad \qquad \left. -\, \tilde{S}(t) \tilde{S}(t - \tau) \tilde{\rho}_{S}(t - \tau)\right] + \text{h.c.}
\end{align}
Here, the system operator and density matrix are written in the interaction picture as $\tilde{S}(t) = e^{i\hat{H}'_St} \hat{S} e^{-i\hat{H}'_St}$ and $\tilde{\rho}_S(t) = e^{i\hat{H}'_St} \hat{\rho}_S e^{-i\hat{H}'_St}$. Note that we have redefined the system Hamiltonian to incorporate the mean-field contribution
\begin{equation}
\label{eq:system_shift}
    \hat{H}_S'= \hat{H}_S + \kappa \braket{\hat{B}}_0 \hat{S},
\end{equation}
which can be done without loss of generality. Eq.~\eqref{eq:weak_coupling} features the bath correlation function (BCF) $C(\tau)$, defined by
\begin{equation}
\label{eq:BCF}
	C(\tau) = \braket{\tilde{B}(\tau) \tilde{B}(0)}_0 - \braket{\hat{B}}_0^2.
\end{equation}

To proceed, we plug the ETH ansatz~\eqref{eq:ETH_ansatz} into the definition of the BCF in Eq.~\eqref{eq:BCF}, finding
\begin{equation}
\label{BCF_2}
	C(\tau) =\sum_{m:m \neq n} e^{-i\omega_{nm}\tau -S(E_{nm})}|f(E_{nm}, \omega_{nm})|^2 |R_{nm}|^2.
\end{equation}
Eq.~\eqref{BCF_2} contains smooth functions from the ETH ansatz and a pseudorandom matrix term which varies erratically. Using the smoothness of these functions combined with small mean level spacing of the spectrum, we can average over this pseudorandom matrix term. In particular, we can define an energy window over which the functions $S(E)$ and $|f(E, \omega)|^2$ do not vary. The mean level spacing is proportional to the inverse density of states, $\delta E \propto [\Omega(E)]^{-1}$. From standard statistical mechanics, the density of states can be expressed in terms of the Boltzmann entropy as $\Omega(E) = e^{S(E)}$. The entropy is extensive, so the level spacing will decrease exponentially in system size, so we expect to find a large number of states within this energy window. For sufficiently long times the phase factor in the BCF could oscillate quickly within this energy window so we require $(t - t') \delta E \ll 1$ which we expect to be true for physically relevant timescales. As a result, the pseudorandom matrix term can be replaced with its average over a small energy and frequency window $\overline{|R_{nm}|^2} = 1$, yielding
\begin{equation}
	\label{BCF_3}
	C(\tau) = \sum_{m: m \neq n} e^{-i\omega_{nm} \tau - S(E_{nm})} |f(E_{nm}, \omega_{nm})|^2.
\end{equation}

Now that the pseudorandom matrix term is removed, we can replace the discrete sum in Eq.~\eqref{BCF_3} with an integral by inserting the density of states, $e^{S(E)}$, to ensure the energy levels are correctly distributed:
\begin{equation}
	\sum_{m:m \neq n} \rightarrow \int dE e^{S(E)} = \int^{\text{max}(\omega)}_{\text{min}(\omega)} d\omega e^{S(E_n + \omega)}.
\end{equation}
The spectral function is known to decay to zero after some frequency scale $\Delta$ \cite{srednicki_approach_1999}. If bath energy $E_{n}$ is sufficiently high and the frequency scale $\Delta$ is sufficiently small, then we can take the bounds of the integral, $\text{max}(\omega)$ and $\text{min}(\omega)$, to be positive and negative infinity, respectively. This restricts the validity of our master equation to sufficiently high temperature initial bath states. Replacing the sum and changing variables to frequency, the BCF is given by
\begin{equation}
	\label{BCF_4}
	C(\tau) = \int^{\infty}_{-\infty} d\omega e^{S(E_{n}+\omega) - S(E_{n} + \frac{\omega}{2})-i\omega \tau} |f(E_{n} + \frac{\omega}{2}, \omega)|^2.
\end{equation}
It is known that the frequency scale $\Delta$ is subextensive \cite{khatami_fluctuation-dissipation_2013} so we can assume that for sufficiently large system size, $\Delta$ will be much smaller then the total bandwidth of the spectrum. The Boltzmann entropy and spectral function are known to be slowly varying in energy \cite{dalessio_quantum_2016}, so we can perform a Taylor expansion of these functions around $E_n$
\begin{equation}
	\begin{aligned}
		&S(E_n + \omega) - S(E_n + \omega/2) \approx \frac{\omega}{2} S'(E_n) + \frac{3\omega^2}{8}S''(E_n),\\
		&|f(E_n + \omega/2, \omega)|^2 \approx |f(E_n, \omega)|^2 + \frac{\omega}{2} \frac{\partial|f(E_n,\omega)|^2}{\partial E}.
	\end{aligned}
\end{equation}
Using the definition of temperature and heat capacity given by 
\begin{equation}
\label{eq:temp_heat}
    \beta = \frac{\partial S}{\partial E} \; \; \text{and} \; \; \mathcal{C} = \frac{\partial E}{\partial T} = - \beta^{-2} \left(\frac{\partial \beta}{\partial E}\right)^{-1},
\end{equation} 
Eq.~\eqref{BCF_4} is written as
\begin{align}
	\label{BCF_finitesize}
		&C(\tau) = \int^{\infty}_{-\infty} d\omega e^{-i \omega \tau} e^{ \frac{\beta \omega}{2} - \frac{3\beta^2 \omega^2}{8\mathcal{C}}} \notag
  \\& \qquad \qquad \times \left[|f(E_n, \omega)|^2 + \frac{\omega}{2} \frac{\partial |f(E_n, \omega)|^2}{\partial E}\right].
\end{align}
The spectral function scales as $|f(E,\omega)|^2 \propto \mathcal{O}(1)$ with system size and so its derivative with energy scales as $\frac{\partial |f(E_n, \omega)|^2}{\partial E} \propto \mathcal{O}(\frac{1}{L})$ due to extensive energy scaling. Additionally, the heat capacity scales extensively $\mathcal{C} \propto \mathcal{O}(L)$. Taking only the leading order terms in system size, we get the following equation for the BCF
\begin{equation}
	\label{BCF_ETH}
	C(\tau) = \int^{\infty}_{-\infty} d\omega e^{-i \omega \tau} e^{\beta \omega/2} |f(E_n, \omega)|^2.
\end{equation}
From the form of the BCF shown in Eq.~\eqref{BCF_ETH}, we have explicitly determined the Fourier transform of the BCF in terms of the spectral function in the ETH.

The next step in the derivation is to use the Markov approximation. This involves taking $\tilde{\rho}_S(t - \tau) \approx \tilde{\rho}_S(t)$, then taking the upper integration limit in Eq.~\eqref{eq:weak_coupling} to infinity, yielding
\begin{align}
\label{eq:Redfield_equation}
    \frac{\partial \tilde{\rho}_{S}}{\partial t} & = \kappa^2\int^{\infty}_{0} d\tau\, C(\tau) \left[ \tilde{S}(t - \tau) \tilde{\rho}_{S}(t)\tilde{S}(t) \right. \notag \\
    & \qquad \qquad \left. -\, \tilde{S}(t) \tilde{S}(t - \tau) \tilde{\rho}_{S}(t)\right] + \text{h.c.}.
\end{align}
Both of these steps are justified in the weak coupling limit when the correlations in the bath decay sufficiently quickly compared to the relaxation dynamics of the system. Specifically, this approximation requires the timescale of decay of the BCF, $\tau_B$, to be small compared to the relaxation timescale, $\tau_R$, over which $\tilde{\rho}_S(t)$ changes appreciably ($\tau_B$ and $\tau_R$ are precisely defined below). The first step is obtained by applying the Taylor series expansion $\tilde{\rho}_S(t - \tau) \approx \tilde{\rho}_S(t) - \tau\frac{\partial\tilde{\rho}_S(t)}{\partial t}$. We can insert this into Eq.~\eqref{eq:Redfield_equation} and show that terms containing $\tilde{\rho}_S(t - \tau)$ are of order $\mathcal{O}(\kappa^4)$ which can be neglected due to weak coupling. Similarly, the quick decay of the correlation function allows us to send the upper integration limit to infinity as the long time contribution of the integral will be negligible. We now show that the validity of the Markov approximation, which depends on the details of the BCF in Eq.~\eqref{BCF_ETH}, can be justified using properties of the ETH spectral function $f(E,\omega)$.

As discussed in Sec.~\ref{sec:set_up}, the spectral function has two crucial properties: it is a smooth function of energy and frequency and it decays to zero at least exponentially in $\omega$ above a frequency scale $\Delta$. The smoothness of this function has been used to show that the BCF asymptotically decays to zero and has been proven to be a sufficient condition for a well-defined weak-coupling limit \cite{parra-murillo_open_2021}. Additionally, the bandwidth of a function and its Fourier transform are inversely proportional to each other \cite{gabor_theory_1947}. Therefore the frequency bandwidth of the spectral function $\Delta$ sets the timescale of decay of the BCF,
\begin{equation}
   \tau_B \propto \Delta^{-1}.
\end{equation}
We note that this frequency scale may in general be temperature dependent; however, this temperature dependence appears to be negligible in our numerical calculations presented in Sec.~\ref{Numerical Results}.

Finally, we make the secular approximation which requires that the timescale of coherent dynamics generated by $\hat{H}_S'$, denoted $\tau_S$, is short compared to the relaxation time $\tau_R$. To formalise this, we rewrite the master equation in the basis of the system energy eigenstates $\hat{H}'_S |\epsilon_{k}\rangle = \epsilon_k |\epsilon_k\rangle$ by decomposing the system observable $S$ into lowering operators, defined by
\begin{equation}
\label{eq:sys_ops_energy}
    \hat{S}(\omega) = \sum_{k,l:\epsilon_l - \epsilon_k = \omega} \braket{\epsilon_k|\hat{S}|\epsilon_l} \ket{\epsilon_k}\bra{\epsilon_l} = \hat{S}^\dagger(-\omega),
\end{equation}
such that $[\hat{H}_S',\hat{S}(\omega)] = -\omega \hat{S}(\omega)$, i.e.~$\hat{S}(\omega)$ acts to decrease the system energy by an amount $\omega$, while $\hat{S}^\dagger(\omega)$ increases it by $\omega$. It follows that 
\begin{equation}
    \label{S_eigenop}
    \tilde{S}(t) = \sum_\omega e^{-i\omega t}\hat{S}(\omega),
\end{equation}
where the sum extends over all transition frequencies (energy gaps) $\omega$, i.e. differences between energy eigenvalues of $\hat{H}_S'$ of the form $\omega = \epsilon_k - \epsilon_l$. Plugging this into Eq.~\eqref{eq:Redfield_equation} yields
\begin{align}
\label{EOM3}
\frac{\partial \tilde{\rho}_{S}(t)}{\partial t} & = \sum_{\omega, \omega'} \Gamma(\omega)e^{i(\omega' - \omega)t}  \\
   &\times \left[ \hat{S}(\omega) \tilde{\rho}_{S}(t) \hat{S}^{\dagger} (\omega')- \hat{S}^{\dagger}(\omega') \hat{S}(\omega) \tilde{\rho}_{S}(t) \right] + \text{h.c.},\notag
\end{align}
where we define the Laplace transform of the BCF
\begin{equation}
   \Gamma(\omega) = \kappa^2 \int^{\infty}_{0} d \tau e^{i\omega \tau} C(\tau).
\end{equation}
Assuming that all oscillatory phase factors vary rapidly compared to the relaxation timescale $\tau_R$, we can make the secular approximation by neglecting all terms in Eq.~\eqref{EOM3} with $\omega\neq \omega'$. Validity of the secular approximation thus requires
\begin{equation}
\label{secular_valid}
    \tau_R \gg \tau_S = \max\left[|\omega-\omega'|^{-1}\right],
\end{equation}
where the maximisation is performed over all distinct pairs of frequencies $\omega\neq \omega'$.

Returning to the Schr\"odinger picture, the dynamics of the system is described by the Lindblad master equation
\begin{equation}
    \label{Lindblad}
    \frac{\partial \hat{\rho}_S}{\partial t} = -i[\hat{H}_S + \kappa\langle \hat{B}\rangle_0 \hat{S}, \hat{\rho}_S] + \sum_\omega \gamma(\omega) \mathcal{D}[\hat{S}(\omega)]\hat{\rho}_S,
\end{equation}
where $\mathcal{D}[\hat{A}]\bullet = \hat{A}\bullet \hat{A}^\dagger - \tfrac{1}{2}\{\hat{A}^\dagger \hat{A},\bullet\}$ denotes a dissipator, with transition rates given by
\begin{equation}
    \label{eq:gamma_def}
    \gamma(\omega) =  \kappa^2\int_{-\infty}^\infty d\tau\, e^{i\omega \tau} C(\tau).
\end{equation}
In writing Eq.~\eqref{Lindblad}, we have ignored the bath-induced renormalisation of the system energy levels (Lamb shift) proportional to ${\textrm Im}[\Gamma(\omega)]$. This is usually negligibly small and has been shown to actually reduce the accuracy of the master equation in some cases~\cite{strasberg_fermionic_2018_real, correa_potential_2024_real}.

The Markov approximation is self-consistent if the BCF~\eqref{eq:BCF} decays to zero on a timescale $\tau_B$ that is small compared to the speed of relaxation of the system. The timescale for the relaxation dynamics is given by 
\begin{equation}
    \tau_{R} \propto \text{min}\left[\gamma(\omega)\right ]^{-1},
\end{equation}
so the Markov approximation requires
\begin{equation}
    \label{eq:Markov_valid}
    \tau_B \propto \Delta^{-1} \ll \tau_R = \text{min}\left[\gamma(\omega)\right ]^{-1}.
\end{equation}
As we have discussed, Markovian dynamics emerges on the timescale given by $\Delta^{-1}$. When the inequality condition in Eq.~\eqref{eq:Markov_valid} is satisfied, the Markovian master equation will describe the system evolution as $\tau_{B}$ is too small a time for it to impact its dynamics.

The transition rate $\gamma(\omega)$, describing the ability of the system to exchange energy $\omega$ with the bath, is now given by Eq.~\eqref{eq:gamma_def}. Using Eq.~\eqref{BCF_ETH} we find that this can be written explicitly in terms of the spectral function
\begin{equation}
\label{eq:gam_spec}
\gamma(\omega) = 2\pi \kappa^2 e^{\beta \omega/2}|f(E_n,\omega)|^2.
\end{equation}
We note the energy dependence of the transition rate is in the spectral function and in the temperature $\beta = \beta(E_n)$. An important symmetry property emerges from Eq.~\eqref{eq:gam_spec} known as local detailed balance \cite{strasberg_OQS_book},
\begin{equation}
   \gamma(\omega) = e^{\beta\omega}\gamma(-\omega).
\end{equation}
This property ensures that the steady state of the dynamics is the thermal state and it allows to us construct a consistent nonequilibrium thermodynamics if the system is driven or in contact with multiple baths. We emphasise that this condition arises from the assumption that the bath is chaotic rather then being an imposed condition \cite{rivas_open_2012_real, breuer_theory_2007_real, PRXQuantum.5.020202, strasberg_OQS_book}. By including the higher order terms in Eq.~\eqref{BCF_finitesize}, we find that decay rate is given by
\begin{equation}
   \gamma(\omega) = 2\pi \kappa^2 e^{ \frac{\beta \omega}{2} - \frac{3\beta^2 \omega^2}{8\mathcal{C}}} \left[|f(E_n, \omega)|^2 + \frac{\omega}{2} \frac{\partial |f(E_n, \omega)|^2}{\partial E}\right],
\end{equation}
where $\mathcal{C}$ is the heat capacity [Eq.~\eqref{eq:temp_heat}].

From these calculations, we have obtained the main result of this paper which is the Born-Markov master equation shown in Eq.~\eqref{Lindblad}, which has been derived for a pure bath state. We have used three primary assumptions to obtain this result. The first and second assumptions are that the bath is prepared in a single eigenstate of the bath Hamiltonian which ensured stationarity and that the system and bath are weakly coupled which allowed us to use Eq.~\eqref{eq:weak_coupling}. The third assumption was that the bath Hamiltonian is chaotic and obeys the ETH. This allowed us to rewrite the BCF in terms of the spectral function [Eq.~\eqref{BCF_ETH}] which ensures the validity of the Markov approximation. This rewriting is exact in the thermodynamic limit and remains valid in large but finite systems at timescales less then the timescale of Poincare recurrences, which are expected to grow as $\tau \propto \mathcal{O}(\exp{[\exp{[L]}]})$ for bath size $L$. 

A further convenient assumption made in this section is the the separation of timescales in Ineq.~\eqref{secular_valid}, which allowed us to invoke the secular approximation to obtain a master equation in Lindblad form. Note, however, that this approximation is not essential: other derivations of a Lindblad equation, resting on different assumptions about the system dynamics, have been widely discussed in the recent literature~\cite{Cresser2017,Farina2019,Nathan2020,Potts2021,Schnell2024}. Nevertheless, all of these derivations ultimately rely on the Markov approximation, which we have shown can be justified from the ETH assuming a pure stationary bath preparation. In the next section, we relax the assumption that the bath is prepared in a single eigenstate and use the ETH to show that the weak-coupling, Lindblad master equation will hold for more general bath preparations.

\subsection{General Bath Preparations}
\label{sec:general_state}

Unlike the single eigenstate case, a generic bath state preparation will not be stationary. The conditions in Eq.~\eqref{stationarity_cond} will then not be satisfied, so the time-convolutionless master equation~\eqref{eq:weak_coupling} cannot be used to describe the system dynamics even at weak system-bath coupling. As a result, our derivation of the Lindblad equation in Sec.~\ref{sec:energy_eigenstate} will generally not be justified.  Nevertheless, here we show that the vast majority of pure states appear approximately stationary for one- and two-point functions of local operators, so that Eq.~\eqref{eq:weak_coupling} remains valid. We prove this result for a class of pure states, called typical microcanonical pure states, which are a random superposition of energy eigenstates $|n\rangle \in \mathcal{E}$ with $\mathcal{E}$ defined as the ensemble of eigenstates that have energies in a microcanonical window $E_n  \in [E - \Delta E/2, E + \Delta E/2]$ and have subextensive fluctuations, $\Delta E/E\propto \mathcal{O}(L^{-1/2})$.

We use Levy's lemma \cite{Milman_typicality_book, Popescu_2006} to show that the expectation value of the bath operator, $\langle \tilde{B}(t)\rangle_0 = \langle\psi_{B}|\tilde{B}(t)|\psi_{B}\rangle$,  when computed with respect to a typical microcanonical pure state $|\psi_{B}\rangle$, has an exponentially suppressed probability of being outside a distance $\epsilon$ from the microcanonical expectation value 
\begin{equation}
\label{eq:levy_exp}
    \text{Pr}\left(|\langle \tilde{B}(t)\rangle_0 - \langle B\rangle_{MC}| > \epsilon \right)  \leq 2 \exp\left[-\frac{d_{\mathcal{E}}\epsilon^2}{18\pi^3\|\hat{B}\|^2}\right].
\end{equation}
Here, $\langle B\rangle_{MC}$ is the microcanonical expectation value, $d_{\mathcal{E}}$ is the dimension of the microcanonical window $\mathcal{E}$, and $\|\hat{B}\|$ is the operator norm (Schatten $\infty$-norm) of the bath operator. Under the assumption that the microcanonical window has a sufficiently narrow energy range, we can assume that $\langle B\rangle_{MC} \approx B(E)$, with $B(E)$ the smooth diagonal function from the ETH. From Levy's lemma, we can also derive a bound for the BCF given by
\begin{equation}
\label{eq:levy_BCF}
    \text{Pr}\left(|C(t, t') - \langle C(\tau)\rangle_{\rm MC}| > \epsilon\right) \leq  4\exp\left[-\frac{ d_{\mathcal{E}}\epsilon^2}{72\|\hat{B}\|^4}\right],
\end{equation}
where $C(t,t')$ and $\langle C(\tau)\rangle_{\rm MC}$ are the BCFs computed with respect to the typical microcanonical pure state $|\psi_B\rangle$ and the microcanonical ensemble, respectively. In App.~\ref{Appendix:typicality}, we show that
\begin{equation}
\label{eq:BCF_ETH_general_state}
    \langle C(\tau)\rangle_{\rm MC} = \int^{\infty}_{-\infty} d\omega e^{-i \omega \tau} e^{\beta \omega/2} |f(E, \omega)|^2.
\end{equation} 
with $\beta$ the inverse temperature of the microcanonical ensemble defined in Eq.~\eqref{eq:temp_heat}. This microcanonical BCF in Eq.~\eqref{eq:BCF_ETH_general_state} is identical to the BCF found in Eq.~\eqref{BCF_ETH} provided the energy window is sufficiently small (see App. \ref{Appendix:typicality} for details). The microcanonical correlation function above will decay to zero at long times, justifying the Markov approximation. Levy's lemma therefore only bounds the most dominant (non-zero) contribution to the integral over the BCF.

For the probabilities in Eqs.~\eqref{eq:levy_exp} and \eqref{eq:levy_BCF} to be small, we require the dimension of the microcanonical window to be large so that the probability of selecting an atypical state is small even when $\epsilon$ is small. The usefulness of these bounds are case sensitive because the choice of a sufficiently small $\epsilon$ depends on the scale set by $\langle B\rangle_{MC}$ and $\langle C(\tau)\rangle_{MC}$, both of which can show complicated model dependent behaviour. For instance, corrections to Eq.~\eqref{eq:BCF_ETH_general_state} are of order $\mathcal{O}(L^{-1})$ in chaotic systems implying that $\epsilon = \mathcal{O}(L^{-2})$ is small on the right scale, which would still be dominated in the exponent by the appearance of $d_E = \mathcal{O}(10^L)$. Additionally, the operator norm $\|\hat{B}\|$ is subextensive due to the locality of the bath operator. Thus, while Levy's lemma does not automatically guarantee that all microcanonical pure states give rise to the same master equation for every system, it clearly indicates the broad applicability of master equations for pure bath states within a microcanonical window~~\cite{bartsch_occurrence_2008,PhysRevLett.102.110403, PhysRevE.97.062129}. This will be particularly true for chaotic systems where fluctuations among different energy eigenstates in a microcanonical window are already strongly suppressed due to the ETH, implying that atypical states also behave like the microcanonical average.

It is natural to ask whether these results require chaotic dynamics, or if master equations can be derived for sufficiently large integrable baths with Haar randomly-sampled pure states. However, physically relevant atypical initial states, which are not initially locally equilibrated, will undergo some nonequilibrium local dynamics. The validity of the master equation derivation for these atypical states requires that under time evolution they become typical or approximately typical and that this occurs on a timescale which is smaller then the timescale associated with the system-environment dynamics. Chaotic Hamiltonians are able generate approximate typicality for atypical initial states after some timescale \cite{PhysRevLett.128.190601, PhysRevE.110.L032203, PhysRevLett.128.180601} ensuring the validity of the master equation; however, we do not expect this to apply to all physically relevant atypical states. Integrable models initially prepared in atypical or low entangled states are expected to evolve to a generalized Gibbs ensemble \cite{vidmar_generalized_2016_real}---an ensemble for which for which master equations may be derived. However, it is also known that in integrable, finite-sized environments large fluctuations around the generalized Gibbs ensemble will occur, potentially affecting the applicability of the master equation for these types of environments. In Sec.~\ref{Numerical Results}, we numerically investigate the dynamics of a system which is locally coupled to chaotic and integrable baths and find decisive differences in their behaviour.

We have now shown that vast majority of typical microcanonical pure states appear stationary for one- and two-point functions when the bath is chaotic and obeys the ETH. This allows us to describe the system dynamics using the time-convolutionless master equation shown in Eq.~\eqref{eq:weak_coupling}. For these states, we have derived an explicit form for the Fourier transform of the BCF, Eq.~\eqref{eq:BCF_ETH_general_state}, which allows us to repeat the same steps that we performed in Sec.~\ref{sec:energy_eigenstate} to derive the Lindblad master equation in Eq.~\eqref{Lindblad}. We will now discuss the master equation derivation for the general interaction Hamiltonian in Eq.~\eqref{V_general}.

\subsection{General Interaction Hamiltonian}
\label{sec:general_int}

In this section, we examine the validity of the master equation for general interaction Hamiltonians as shown in Eq.~\eqref{V_general}. Under the same assumptions leading to Eq.~\eqref{eq:weak_coupling}, the influence of the bath is characterised by a set of two point functions of the form
\begin{equation}
\label{BCF_gen_interaction}
    C_{\mu \nu}(t, t') = \langle \tilde{B}^{\mu}(t) \tilde{B}^{\nu}(t')\rangle_0 - \langle \tilde{B}^{\mu}(t') \rangle_0 \langle \tilde{B}^{\nu}(t')\rangle_0.
\end{equation}
Using a generalization of the ETH, we will find an explicit expression for the Fourier transform of the BCF in terms of smooth functions in energy and frequency. These smooth functions can be transformed so that their properties are the same as the spectral function which appears in the conventional ETH ansatz. This ensures that the system dynamics can be described by a Lindblad master equation and that local detailed balance holds. We prove this result for eigenstate and typical microcanonical pure states. For details of these calcuations, see App. \ref{Appendix:general_interaction}.

Recent extensions of the ETH have allowed spectral functions to be defined for products of distinct operators. Consider operators $\hat{B}^{\mu}$ and $\hat{B}^{\nu}$ whose off-diagonal elements in the basis of the energy eigenstates are described by a pseuodrandom matrix of the form $R^{\mu}_{nm}$ and $R^{\nu}_{nm}$, respectively. In general, these random matrices are not uncorrelated; however, it has been shown that correlations between these matrices can be characterized by a smooth function \cite{noh_numerical_2020, pappalardi_eigenstate_2022, pappalardi2024eigenstatethermalizationfreecumulants},
\begin{equation}
    \overline{R^{\mu}_{nm} R^{\nu}_{mn}} = h^{\mu \nu}(E, \omega).
\end{equation}
Here, $\overline{\bullet}$ denotes averaging over the matrix elements of these operators within a small energy and frequency window.

Using this extension to the ETH ansatz, we show that the BCF in Eq.~\eqref{BCF_gen_interaction} can be written as
\begin{equation}
\begin{aligned}
    &C_{\mu \nu}(\tau) = \int^{\infty}_{-\infty}d \omega e^{-i \omega\tau} e^{\beta \omega/2} F^{\mu \nu}(E_n, \omega).
\end{aligned}
\end{equation}
Above, $E_n$ is the energy of the initial state, taken to be a single eigenstate $|n\rangle$, and the spectral function is defined as $F^{\mu \nu}(E, \omega) = f^{\mu}(E, \omega) h^{\mu \nu}(E, \omega) f^{\nu}(E, -\omega)$. We used the smoothness of $F^{\mu \nu}(E, \omega)$ to replace this sum with an integral and have also assumed that $F^{\mu \nu}(E_n, \omega)$ decays exponentially for positive and negative frequencies after $\Delta^+_{\mu \nu} \ll \max(\omega)$ and $\Delta^-_{\mu \nu} \gg \min(\omega)$ so that the bounds of the integral can be taken to $\pm \infty$. This is justified from the contributions of $f^{\mu}(E, \omega)$ and $f^{\nu}(E, -\omega)$, which both decay exponentially in frequency, and assuming that $h^{\mu \nu}(E, \omega)$ does not grow exponentially for large frequencies.

We then define the following matrix,
\begin{equation}
\begin{aligned}
    &\gamma_{\mu \nu}(E_n, \omega) = \kappa^2\int^{\infty}_{-\infty}d\tau e^{-i\omega \tau} C_{\mu \nu}(\tau) \\
    & \qquad \qquad \;\;\; = 2\pi \kappa^2 e^{\beta \omega/2} F^{\mu \nu}(E_n, \omega),
\end{aligned}
\end{equation}
whose eigenvalues are the transition rates. In App.~\ref{Appendix:general_interaction}, we use the properties of the spectral function to show that the eigenvalues of this matrix are non-negative, ensuring that the dynamics is described by a Lindblad master equation and is completely positive and trace-preserving. This is done by redefining the system and bath operators as $\tilde{K}^{\bar{\mu}}(\tau) = \sum_{\mu} [U^{\dagger}]^{\bar{\mu} \mu} \tilde{S}^{\mu}(\tau)$ and $\tilde{A}^{\bar{\mu}}(\tau) = \sum_{\mu} [U^{\dagger}]^{\bar{\mu} \mu} \tilde{B}^{\mu}(\tau)$, with $U^{\bar{\mu} \mu} \equiv U^{\bar{\mu} \mu}(E_n, \omega)$ the unitary that diagonalizes $F^{\mu \nu}(E_n, \omega)$.

The transition rates are shown to be given by
\begin{equation}
    \gamma_{\bar{\mu}}(E_n, \omega) = 2\pi \kappa^2 e^{\beta \omega/2}|f_{\bar{\mu}}(E_n, \omega)|^2,
\end{equation}
with $f_{\bar{\mu}}(E_n, \omega)$ the spectral function associated with operator $\tilde{A}^{\bar{\mu}}$. These rates are positive, real, and obey local detailed balance due to the symmetry property $|f_{\bar{\mu}}(E, \omega)|^2 = |f_{\bar{\mu}}(E, -\omega)|^2$. Using this result, the Lindblad master equation can be written as follows,
\begin{equation}
    \label{Lindblad_gen_int}
    \frac{\partial \hat{\rho}_S}{\partial t} = -i[\hat{H}'_S, \hat{\rho}_S] + \sum_{\omega, \bar{\mu}} \gamma_{\bar{\mu}}(\omega) \mathcal{D}[\hat{K}^{\bar{\mu}}(\omega)]\hat{\rho}_S.
\end{equation}
The redefined system operators, $\hat{K}^{\bar{\mu}}$ can still interpret these operators as raising and lowering operators due to the commutation relation $[\hat{H}_S',\hat{K}^{\bar{\mu}}(\omega)] = -\omega \hat{K}^{\bar{\mu}}(\omega)$.

We have shown that the ETH ensures consistent nonequilibrium thermodynamics for a local subsystem coupled to a chaotic bath for a variety of initial states and interaction Hamiltonians. This master equation derivation will also hold for a system coupled to multiple chaotic baths ensuring the applicability of the master equation to transport scenarios. In the next section we will numerically verify this master equation description for a chaotic bath Hamiltonian by comparing it to dynamics computed using exact diagonalization. This will be done for a variety of pure initial states and temperatures and we will also investigate the behaviour of an integrable bath where the ETH does not hold.

\section{Numerical Results}
\label{Numerical Results}

\subsection{Many-Body Framework and ETH Data}
\label{Numerical Results:1}

To compute the master equation and exact dynamics, we must first choose a many-body model for our system, bath and interaction. For the system Hamiltonian we consider a single spin given by
\begin{equation}
\label{eq:sys_ham}
    \hat{H}_{S} = \frac{\omega_0}{2}\hat{\sigma}^z_0,
\end{equation}
with $\omega_0$ being system energy splitting, for which we choose the numerical value $\omega_{0} = 1.525$. To test our master equation, we want to compare chaotic and integrable dynamics and so we will consider a bath Hamiltonian that can be tuned between these regimes. Specifically, we choose a mixed-field one-dimensional Ising model with boundary terms to break inversion symmetry:
\begin{equation}
\label{TL_Ising}
    \hat{H}_{B} =  J\sum^{L-1}_{j =1} \hat{\sigma}^z_j \hat{\sigma}_{j+1}^z + \sum^L_{j = 1} \left(h_z \hat{\sigma}^z_j + h_x \hat{\sigma}^x_j\right) + h_1 \hat{\sigma}^z_1 + h_L \hat{\sigma}^z_L.
\end{equation}
For generic parameters this Hamiltonian is chaotic. For our calculations in the chaotic regime, we choose the parameters $(J, h_z, h_x, h_1, h_L) = (1.0, 0.3, 1.1, 0.25, -0.25)$, which are believed to be maximally chaotic \cite{PhysRevX.14.031014}. The Ising model becomes integrable if we set the longitudinal field to zero, the parameters are then $(J, h_{z}, h_{x}, h_{1}, h_{L}) = (1.0, 0.0, 1.1, 0.0, 0.0)$. We use this Hamiltonian with both of these sets of parameters to investigate if the master equation accurately describes the dynamics of a system coupled to a chaotic bath. The system-bath interaction Hamiltonian is given by
\begin{equation}
\label{eq:int_ham}
    \kappa \hat{V} = \kappa \hat{\sigma}^x_0\otimes \hat{\sigma}^x_{1}.
\end{equation}
with $\kappa = 0.15$. Throughout this paper, we consider bath system sizes of $L = 15$. For our numerical simulations, we use QuSpin \cite{phillip_weinberg_quspin_2017} to perform the exact diagonalization.

As a first check, we ensure that the chaotic bath Hamiltonian obeys the ETH. We perform exact diagonalization of both the chaotic and integrable bath Hamiltonians and look at the diagonal elements of the bath operator $\hat{B} = \hat{\sigma}^{x}_{1}$ in the basis of the energy eigenstates. In the inset of Fig. \ref{fig:ETH_data}a, we see exponential collapse of the support of the diagonals as system size increases, measured using the eigenstate-to-eigenstate fluctuations. In contrast, the integrable case, shown in Fig. \ref{fig:ETH_data}b, does not have this exponential collapse of support as these eigenstates do not obey the ETH. We note that these properties of the Ising model are well known from previous studies~\cite{kim_testing_2014}.

We also look at the off-diagonal matrix elements and extract the spectral function. To do this, we first select energy eigenvalues within a narrow energy window. The average energy of this window will correspond to a temperature which can be determined by computing the derivative of the logarithm of the density of states with respect to the energy which can be obtained by exact diagonalization of the bath Hamiltonian. Details on computing temperature from the energy are given in App. \ref{temperature_Appendix}. Within the energy window, we then select matrix elements with frequency within a narrow frequency window. Averaging over matrix elements within these frequency windows will allow us to average over the pseudorandom matrix term using the fact that the variance is given by $\overline{|R_{nm}|^2} = 1$. The frequency window size is chosen to be sufficiently small so that the spectral function is constant within the window, but large enough so that we can average over the pseudorandom matrix term. The spectral function is only defined up to a constant factor which is dependent on the size of the frequency window. We can fix this factor by noting that the following integral of the transition rate should hold using Eq.~\eqref{eq:gamma_def},
\begin{equation}
    \int^{\infty}_{-\infty} d\omega \gamma(\omega)  = 2\pi\kappa^2 \left[\langle \hat{B}^2\rangle_0 - \langle \hat{B}\rangle_0^2\right].
\end{equation}
Then we use Eq.~\eqref{eq:gam_spec} and we can see that the constant factor can be fixed by ensuring the following equality holds,
\begin{equation}
\label{spec_func_normalize}
    \int^{\infty}_{-\infty} d \omega e^{\beta\omega/2}|f(E,\omega)|^2 = \langle \hat{B}^2\rangle_0 - \langle \hat{B}\rangle_0^2.
\end{equation}

In Fig. \ref{fig:ETH_data}c, we have extracted smooth functions of frequency from the off-diagonals for the chaotic bath and for different temperatures. We observe minimal temperature dependence which is a result of the insensitivity of the spectral function to changes in energy. The off-diagonals of the integrable case, shown in Fig. \ref{fig:ETH_data}d, are sparse due to local symmetries of the Hamiltonian. We define a spectral function in this case by averaging over larger frequency bins. 
\begin{figure}[h]
    \centering
    \includegraphics[width=0.47\textwidth]{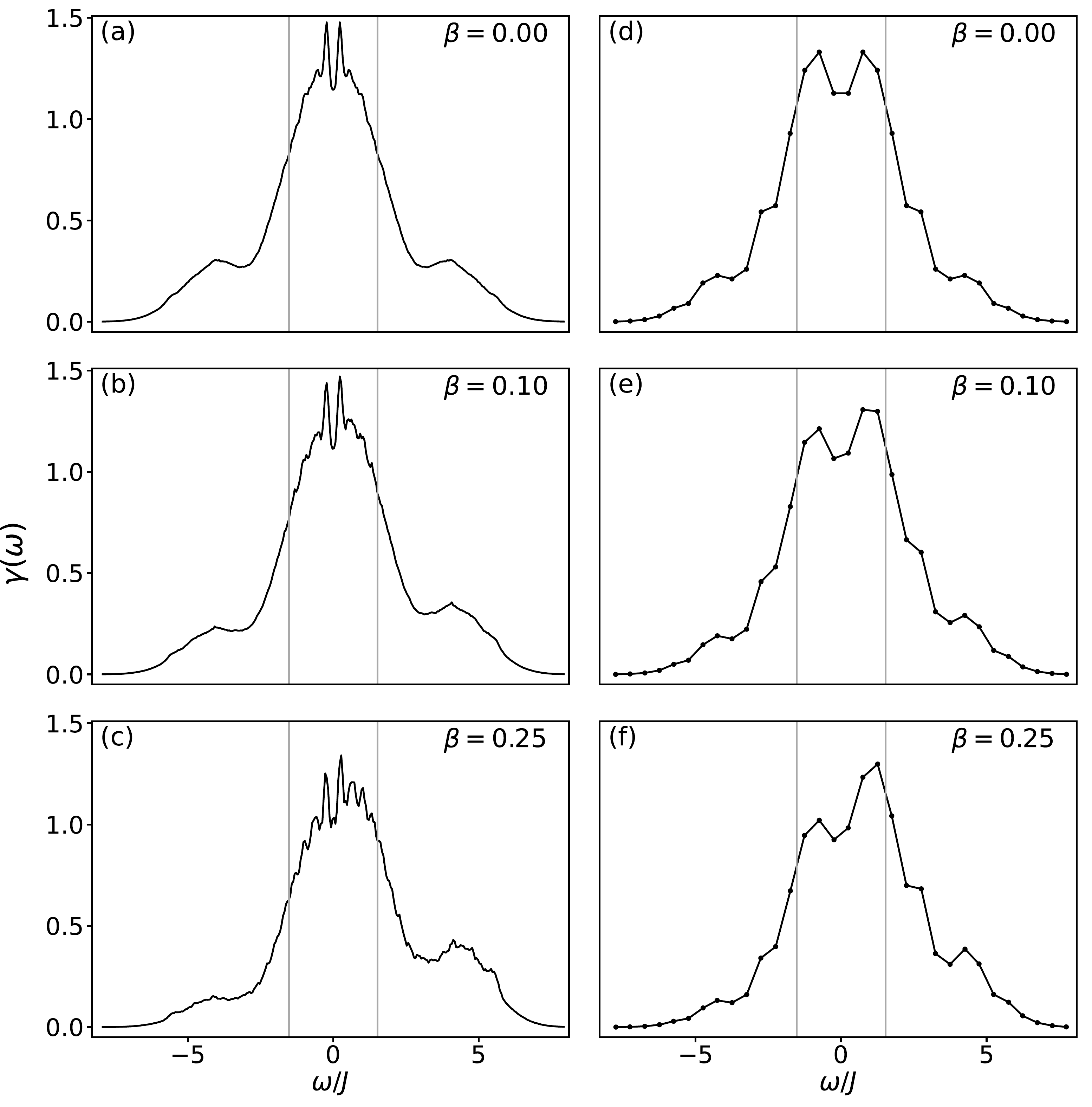}
    \caption{[(a), (b), (c)] the transition rates for the chaotic Ising model as a function of system frequency $\omega$ for different temperatures. These are computed using the relation in Eq.~\eqref{eq:gam_spec} and the ETH data shown in Fig. \ref{fig:ETH_data}. They are smooth functions of frequency as predicted by the ETH. [(d), (e), (f)] the transition rates computed for the integrable Ising model for different temperatures. Larger bins are needed to perform the averaging because the off-diagonals are not described by a smooth function. The grey vertical lines indicate the system frequency $\pm\omega_0$.}
    \label{fig:gamma}
\end{figure}

Using the spectral function and the relation in Eq.~\eqref{eq:gam_spec}, we compute the transition rates. In Figs. \ref{fig:gamma}a, \ref{fig:gamma}b, \ref{fig:gamma}c, the transition rates for the chaotic model are shown for different temperatures and are observed to be smooth functions of frequency. As temperature decreases we observe more noise in the transition rates due to finite-size effects which become more prominent near the edges of the spectrum. This is due to a reduced number of states within the energy window and we expect that for smaller temperatures we would not be able to define a smooth function of the frequencies. We also compute the transition rate for the system coupled to an integrable bath shown in Fig. \ref{fig:ETH_data}. We then normalize these transition rates shown in Eq.~\eqref{spec_func_normalize}. In Figs. \ref{fig:gamma}d, \ref{fig:gamma}e, \ref{fig:gamma}f, the transition rates for the integrable model at different temperatures is computed and we see that, due to the sparseness, we have less data points which are represented by the dots. The grey vertical line indicates the system frequency which we will be using in our numerical calculations. In both cases, moving away from the infinite temperature regime results in an asymmetric transition rate around the $\omega = 0$ axis. This asymmetry arises because decreases in the system energy become entropically favorable, since the Boltzmann entropy of the bath will increase when gaining energy.

\subsection{Dynamics}

We now look at the dynamics of the chaotic and integrable models for different initial bath preparations with average energies corresponding to temperatures $\beta = 0.0, 0.1$ and $0.25$. We consider three kinds of pure initial bath states: energy eigenstates, typical microcanonical (MC) pure states (i.e.~random superpositions of eigenstates within an energy window, as in Sec.~\ref{sec:general_state}), and product states. To generate the typical microcanonical pure states, we take an energy window with $\Delta E = 0.3$ containing approximately 600 eigenstates. This energy window is small enough that the density of states looks constant within it, yet large enough to contain enough states to meaningfully average over~\cite{burke_entropy_2023}. We then select the coefficients $\{c_{1}, c_{2}, ..., c_{n}\}$ randomly from a Gaussian distribution with zero mean and unit variance. Importantly, we do not average over different initial states: we pick a single state for each temperature and perform our calculations for this state. The product states are chosen so that they have an average energy corresponding to the temperatures above. We also consider two different pure initial states for the system to display the dynamics of the populations and coherences,
\begin{equation}
\label{sys_states}
    |\psi_{S}\rangle = |0\rangle \; \; \text{and} \; \; |\psi_{S}\rangle = \frac{1}{2}(|0\rangle + |1\rangle).
\end{equation}

\begin{figure}[t]
\centering
\includegraphics[width=0.47\textwidth]{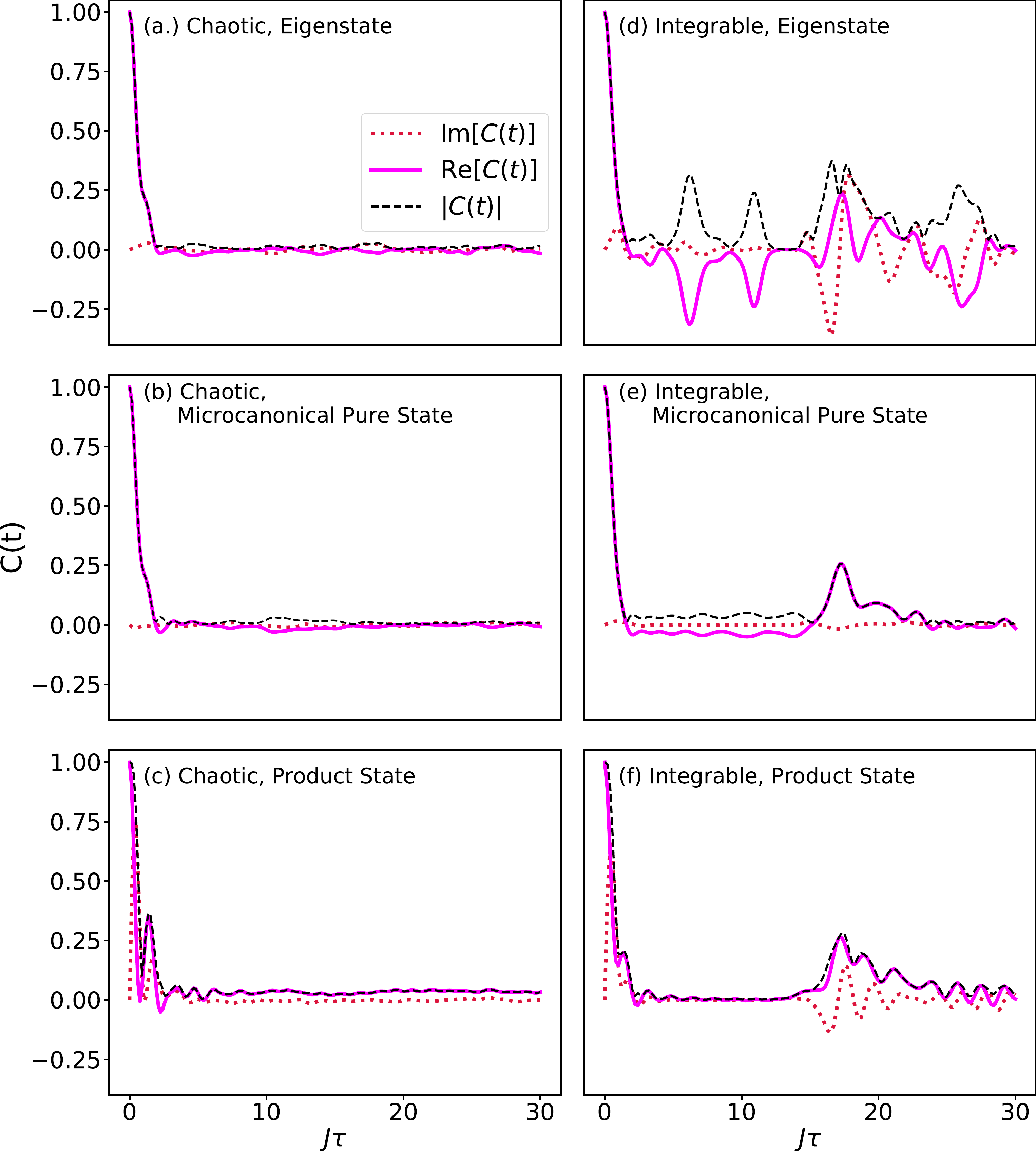}
\caption{BCF computed using exact diagonalization for chaotic [(a), (b), (c)] and integrable [(d), (e), (f)] models with bath size $L = 15$. We consider infinite temperature and initial bath states given by a single eigenstate [(a), (d)], typical MC state [(b), (e)] and product state [(c), (f)].}
\label{fig:BCF}
\end{figure}

First we compute the BCF as a function of time using exact diagonalization for the bath operator given by Eq.~\eqref{eq:int_ham}. The BCF data for the chaotic and integrable model is shown in Fig. \ref{fig:BCF}a, \ref{fig:BCF}b, \ref{fig:BCF}c and Fig. \ref{fig:BCF}d, \ref{fig:BCF}e, \ref{fig:BCF}f, respectively. We only consider the infinite temperature bath state in this instance as we want to show the difference in the long time fluctuations. In the chaotic case, for the single eigenstate and typical MC state we observe an initial decay to zero and small late-time fluctuations which are a result of finite size effects. The product state exhibits comparatively larger oscillations at short times and then decays to zero. The timescale of decay given by the half width at half maximum is approximately $\tau_{B} = 0.5$ which is approximately the inverse of the energy scale of the FFHW of the spectral function in Fig. \ref{fig:ETH_data}c. We expect Poincare recurrences to occur at much longer timescales which will cause fluctuations of the BCF. The integrable case also decays to zero, but it experiences more fluctuations and has a recurrence at $\tau \approx 18$. This is indicative of the reduced effective dimension of integrable models due to the extensive number of conserved quantities that constrain the dynamics. These fluctuations ensure that the dynamics cannot be described by a time homogeneous generator.

We also investigate the possibility that equilibration occurs due to the interaction term breaking integrability of the overall system-bath Hamiltonian. It is known that integrability in interacting many-body models is not robust against perturbations \cite{santos_speck_2020, brenes_eigenstate_2020, brenes_low-frequency_2020} and in App. \ref{level_statistics} we observe that the level spacing statistics of the overall system and environment does not agree with the Wigner-Dyson distribution expected in chaotic systems~\cite{dalessio_quantum_2016}. The exact dynamics is also not described by the mean-force state so it is clear that its behaviour is not fully described by the energy of the initial state. We note that even if this does not describe equilibrium, the size of the perturbation required to break integrability decreases exponentially in system size~\cite{Pandey2020} suggesting that in the thermodynamic limit, the master equation description may hold for an integrable bath Hamiltonian.

\subsubsection{Populations}
\begin{figure}[h]
\centering
\includegraphics[width=0.49\textwidth]{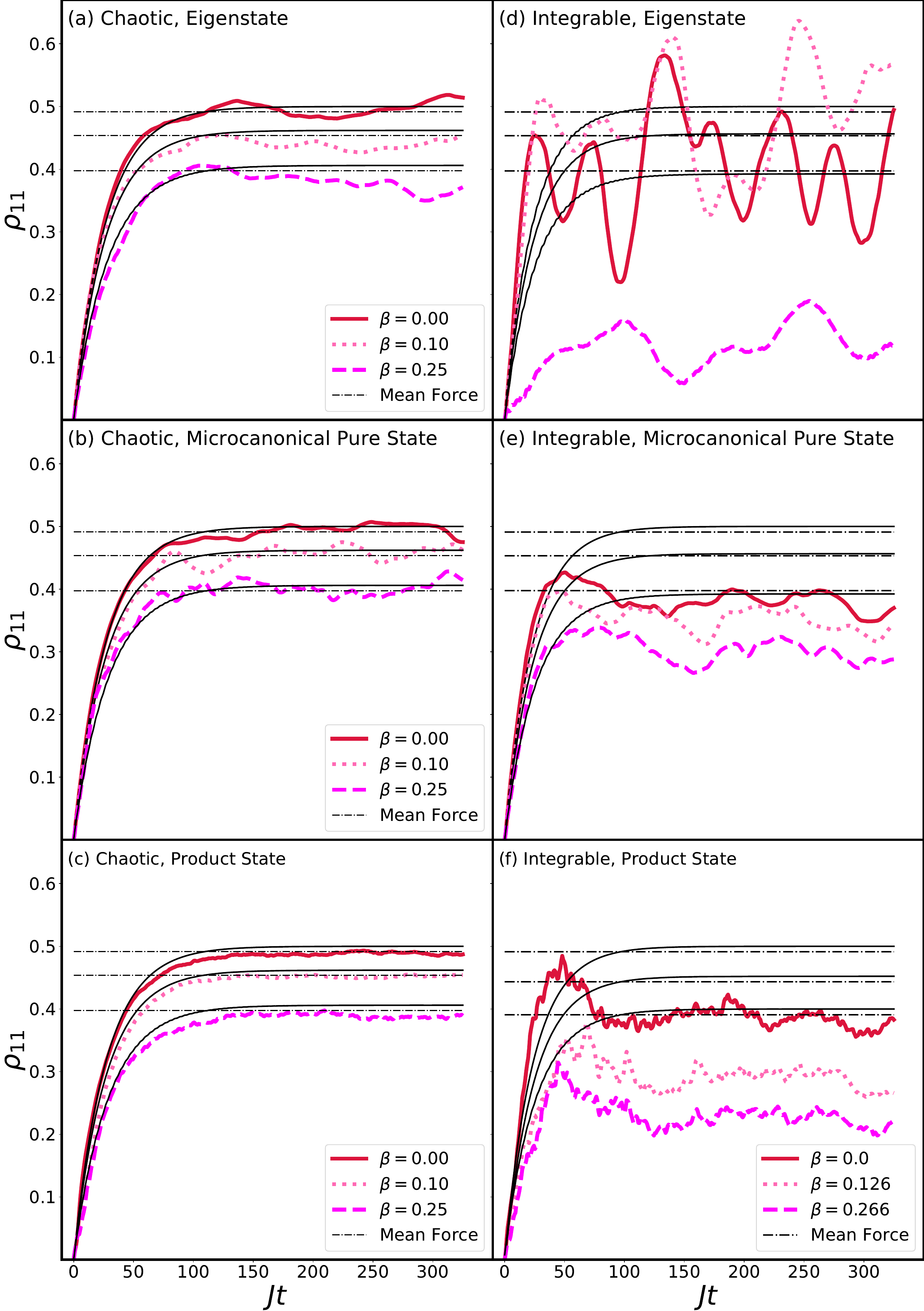}
\caption{Population dynamics computed using exact diagonalization and the master equation. The state of the system is initially prepared in a polarized state as shown in Eq.~\eqref{sys_states} and we consider multiple different temperatures. [(a), (b), (c)] displays a system coupled to a chaotic bath for the eigenstate (a), typical MC (b) and product state (c). We see close agreement between the master equation (black) and exact dynamics for different temperatures with finite size fluctuations affecting the single eigenstates most strongly. Small bias in the long time behaviour is corrected by the mean-force state in Eq.~\eqref{MF}. [(d), (e), (f)] displays dynamics for a system coupled to an integrable bath and we observe much larger fluctuations and stronger bias away from thermal behaviour at long times.}
\label{fig:populations}
\end{figure}

We compute the behaviour of the populations as a function of time for the different bath preparations and for the system prepared in the state $|\psi_{S}\rangle = |0\rangle$. In Figs. \ref{fig:populations}a, \ref{fig:populations}b, \ref{fig:populations}c, we observe the population dynamics for the system coupled to the chaotic bath. We see agreement between the master equation (black) and exact dynamics for all bath preparations considered and for all temperatures. The single eigenstate case suffers most from fluctuations due to finite size effects; we expect that for larger system sizes these fluctuations should decrease. The typical MC and product state dynamics are well described by the master equation, since finite-size effects are averaged out by the contribution from the large number of energy eigenstates which they comprise. We expect that at larger system sizes the results would converge exactly to the master equation description. 

The populations should thermalize exponentially at a rate given by 
\begin{equation}
    \gamma_{pop} = \gamma(\omega_0) + \gamma(-\omega_0) = \gamma(\omega_0) \left[1 + e^{-\beta \omega_0}\right].
\end{equation}
Note, we neglect the contribution proportional to $\gamma(0)$ as it is on the order of $\mathcal{O}(10^{-5})$ times smaller than the other terms and is therefore negligible. In the product state preparation, equilibration of the populations occurs but with a small bias in the late time behaviour. We plot the mean force population (black dashed dotted) which is defined by the following equation \cite{strasberg_OQS_book, trushechkin_open_2022_real}
\begin{equation}
\label{MF}
    \hat{\rho}_{\text{MF}} \propto \text{tr}_{B} \left[ e^{-\beta \hat{H}}\right].
\end{equation}
Here, $H$ is the total Hamiltonian of the system and bath and $\beta$ is the temperature of the full state $|\psi_{SB}\rangle$. We see that the mean-force correction accounts for most of the bias in the saturation of the populations. This suggests that this bias is a result of the coupling between system and bath being too strong, thus inducing correlations which affect the bath and reduce the accuracy of the Born approximation. For sufficiently large systems sizes, we could decrease the coupling constant to reduce this error. However, in finite sized systems the coupling constant must be larger than the mean level spacing of the bath spectrum in order to induce any nontrivial dynamics, so we cannot make it arbitrarily small.

In Figs. \ref{fig:populations}b, \ref{fig:populations}d, \ref{fig:populations}f, we plot the populations for the integrable bath. We observe much larger fluctuations for all bath preparations. As in the chaotic bath, the single eigenstate case suffers most from these fluctuations and is not described by the master equation. For the typical MC and product state preparations, we see less fluctuations and a bias to smaller temperature populations in the long time behaviour compared to the prediction of the master equation. This bias in the long-time behaviour is not described by the mean force state; instead, it is likely a result of a restriction on transitions in the bath due to conserved charges. Note that we used product states with temperature closest to the eigenstate and typical microcanonical pure state; however, we could not obtain the same temperatures as we have been using in the other cases. 

\subsubsection{Decoherence}
\begin{figure}[t]
\centering
\includegraphics[width=0.49\textwidth]{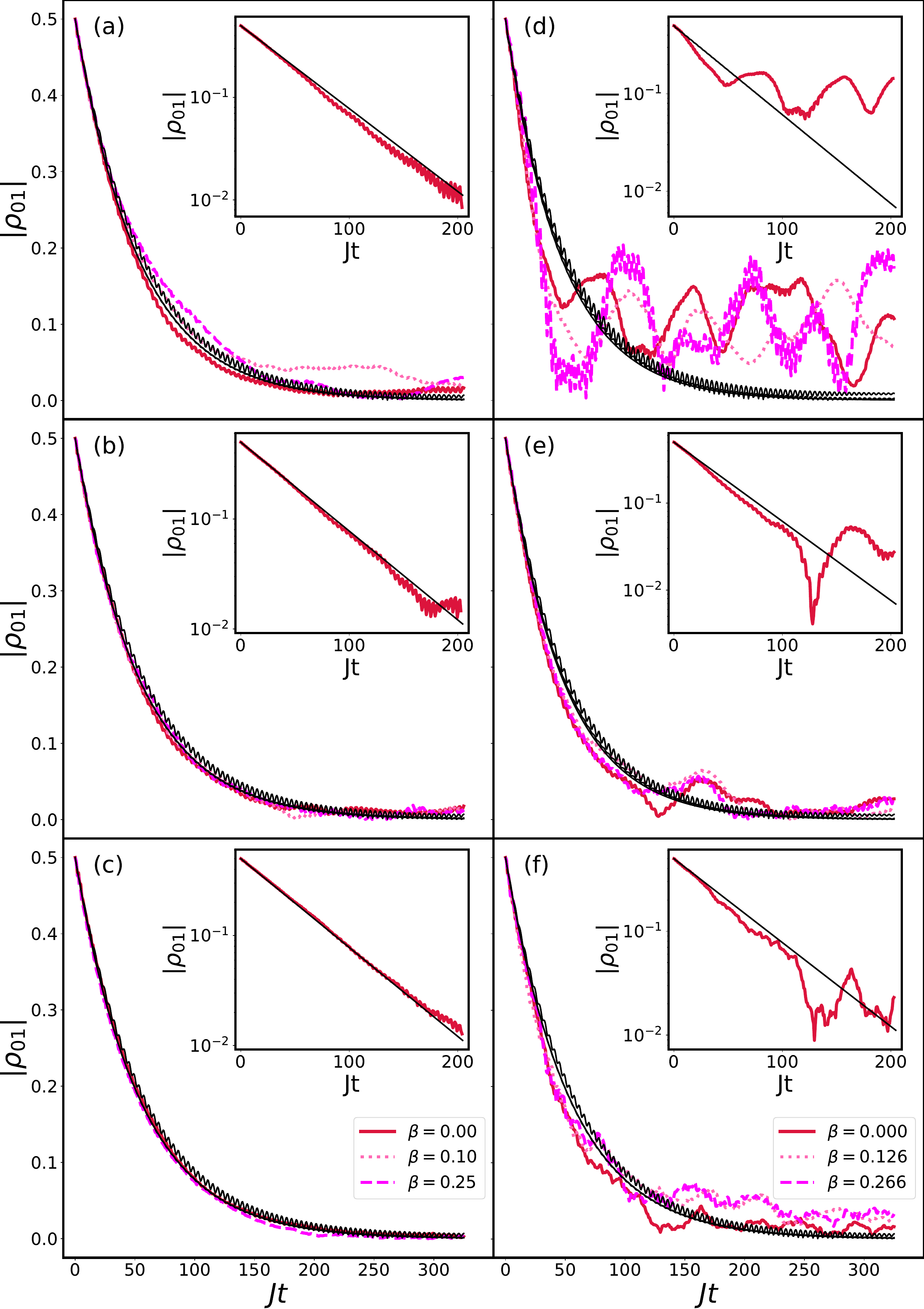}
\caption{Coherence dynamics computed using the exact diagonalization and the master equation. The state of the system is initially prepared in an unpolarized state as shown in Eq.~\eqref{sys_states} and we consider multiple different temperatures as we have done in Fig. \ref{fig:populations}. [(a), (b), (c)] displays a system coupled to a chaotic bath for the eigenstate (a), typical MC (b) and product state (c). The dynamics has very little temperature dependence and the exact dynamics is well described by the master equation, albeit with significant finite-size effects in the single eigenstate case. [(d), (e), (f)] displays dynamics for a system coupled to an integrable bath and we observe much larger fluctuations and a bias to larger decay rates for the single eigenstate and product state case. Insets: the coherence for the infinite temperature bath show approximately exponential decay, but superimposed with prominent fluctuations in the integrable case.}
\label{fig:coh}
\end{figure}

Next, we look at the decay of coherences in a system prepared in the unpolarized state $|\psi_{S}\rangle = \frac{1}{\sqrt{2}}(|0\rangle + |1\rangle)$. In Figs. \ref{fig:coh}a, \ref{fig:coh}b, \ref{fig:coh}c we observe the decay of the off-diagonals of the system density matrix as a function of time for the system coupled to the chaotic bath. As in the previous case, we see close agreement for all types of initial states considered with finite size effects impacting the single eigenstate case the most. In the insets we see that decoherence occurs with a rate given by
\begin{equation}
    \gamma_{coh} = \frac{\gamma(\omega_0) + \gamma(-\omega_0)}{2} = \gamma(\omega_0) \left[\frac{1 + e^{-\beta \omega_0}}{2}\right],
\end{equation}
which is predicted by the master equation. 

In Figs. \ref{fig:coh}d, \ref{fig:coh}e, \ref{fig:coh}f we see the integrable bath case and observe strong fluctuations in the exact dynamics which is not captured by the master equation. The single eigenstate case has a decay rate which is not captured by the master equation as it suffers from large fluctuations. The decay of the typical MC bath state and the product state is well described by the master equation and shows less fluctuations than the single eigenstate. This is again expected due to the effective averaging over many energy eigenstates within the superposition. The result for the typical MC state is also consistent with our typicality argument, which is independent of the chaotic or integrable nature of the bath dynamics. Despite the better agreement, the master equation is still not as accurate as in the chaotic case, especially at longer times. 

\subsection{Finite-Size Scaling}
We analyse the convergence of the exact dynamics to the master equation dynamics as system size increases. We perform this analysis for a single eigenstate and pure microcanonical bath preparation, both having infinite temperature. We compute the system density matrix for bath sizes $L = 6, 8, 10, 12, 14$ using exact dynamics, $\hat{\rho}^{\text{exact}}_{S}(t)$. We then compute the trace distance between the exact result and the result obtained from the master equation, $T\left(\hat{\rho}^{\text{exact}}_{S}, \hat{\rho}^{\text{master}}_{S}\right) = \frac{1}{2}\text{tr}\left[\sqrt{(\hat{\rho}^{\text{exact}}_{S} - \hat{\rho}^{\text{master}}_{S})^{\dagger} (\hat{\rho}^{\text{exact}}_{S} - \hat{\rho}^{\text{master}}_{S})}\right]$. Finally, we take the average of this trace distance up to final time $t'$,
\begin{equation}
    \langle T\rangle_t = \frac{1}{t'}\int^{t'}_0 dt T\left(\hat{\rho}^{\text{exact}}_S(t), \hat{\rho}^{\text{master}}_S(t))\right).
\end{equation}

\begin{figure}[h]
\centering
\includegraphics[width=0.475\textwidth]{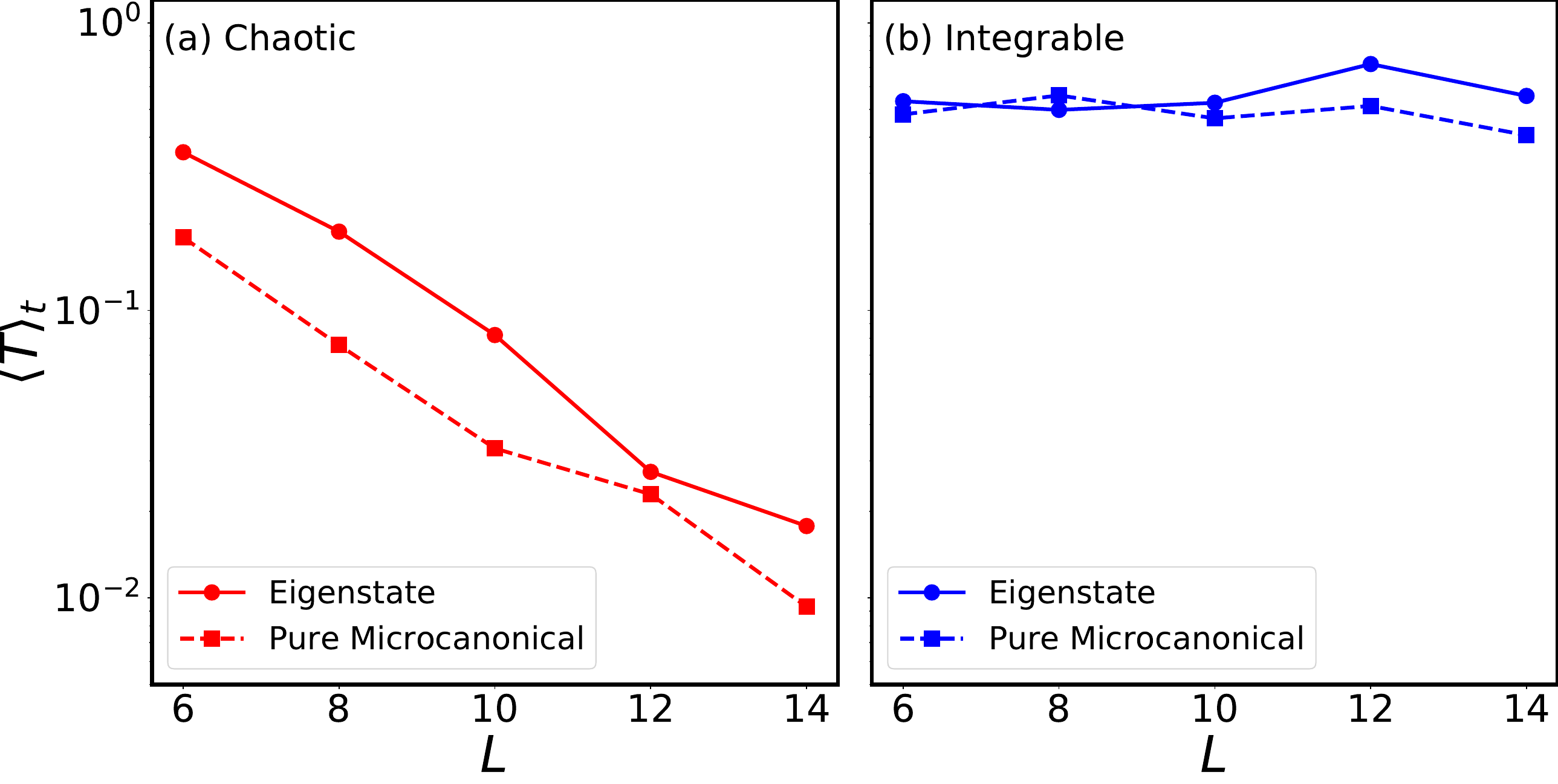}
\caption{System size scaling analysis of the time-averaged distance between the system density matrix computed using exact and master equation dynamics. This is performed for chaotic (a) and integrable (b) baths with two initial bath states: single eigenstate and a pure microcanonical state both with infinite temperatures. The system is prepared initially in the state $|\psi_S\rangle = |0\rangle$. We observe exponential decay in system size of this distance for two different initial states in the chaotic case.}
\label{fig:dynamics_sys_scaling}
\end{figure}

In Fig.~\ref{fig:dynamics_sys_scaling}, we compute this time-averaged trace distance for $t' = 325$ and for initial system preparation $\rho_{S}(0) = |0\rangle \langle 0|$. For the chaotic bath case, Fig.~\ref{fig:dynamics_sys_scaling}a, exponential decay of the time-averaged trace distance as a function of system size is seen. We believe this is strong evidence that, in the thermodynamic limit, the exact dynamics will be described the master equation.

The integrable bath case, Fig.~\ref{fig:dynamics_sys_scaling}b displays no decay of the trace distance for increased system size. However, we cannot necessarily conclude that the integrable bath case will never be described by the master equation, it may be the case that a sufficiently large, integrable environment will cause decaying correlation functions which would allow for an approximate master equation description, but at these system sizes it is not observed.

\section{Discussion and Conclusion}

In this paper, we have established chaos and the ETH as primary requirements for the applicability of the weak-coupling Lindblad master equation to a system coupled to a finite-sized bath prepared in a pure state. The pure states considered appear equilibrated for local observables, allowing us to explicitly obtain the Fourier transform of the BCF shown in Eq.~\eqref{BCF_ETH}. Using the properties of the spectral function in the ETH, we show that the BCF decays on a timescale set by the inverse bandwidth of the spectral function in frequency. For sufficiently weak coupling, the BCF decays on a timescale that is much shorter than the timescale of interaction between the system and bath, allowing us to derive a weak-coupling, Lindblad master equation. An expression for the transition rates in terms of the spectral function in the ETH is given in Eq.~\eqref{eq:gam_spec} and can be used to show local detailed balance. This master equation description will hold for the vast majority of pure states with sufficiently high energy and subextensive energy fluctuations. The bath is assumed to be finite, but sufficiently large such that the energy spectrum appears smooth. Our master equation generalises straightforwardly to an open system coupled to multiple chaotic baths, so long as the aforementioned conditions hold for each bath in isolation.

At weak coupling, the OQS dynamics depends only on the (two-time) BCF and it is instructive to compare it with the widely used Caldeira-Leggett model in OQS theory. For large baths, using Eq.~\eqref{BCF_ETH} and with the notation of \cite{strasberg_OQS_book}, we find that a Caldeira-Leggett model with spectral density
\begin{equation}
\label{eq:spec_density}
    J(\omega) = 2\pi \sinh(\beta\omega/2) |f(E,\omega)|^2
\end{equation}
gives rise to the same dynamics as the ETH master equation. Indeed, Eq.~\eqref{eq:spec_density} is nothing but the dissipative response function, usually denoted by $\chi''(\omega)$ in linear-response theory~\cite{Forster2018}, which quantifies the characteristic rate of energy absorption by the bath at each frequency. The main benefit of our derivation is thus to show that neither an integrable Hamiltonian nor a Gibbs state for the bath are needed in deriving a quantum master equation with a consistent nonequilibrium thermodynamics. Moreover, notice that the spectral density will in general explicitly depend on the bath temperature via Eq.~\eqref{eq:spec_density}. Once we leave the weak-coupling regime, higher order correlation functions of the bath will become important and explicit differences from a Gaussian (Caldeira-Leggett) bath will appear. Those could be handled within the full ETH framework \cite{foini_eigenstate_2019, pappalardi_eigenstate_2022, brenes_out--time-order_2021, pappalardi2024eigenstatethermalizationfreecumulants} and this constitutes an interesting direction for future research.

We numerically verified our theory by comparing the master equation to dynamics calculated using exact diagonalization. For the chaotic bath, we found close agreement between the exact and master equation dynamics for eigenstate, typical microcanonical pure state, and product state bath preparations at a variety of temperatures. In contrast, the integrable bath shows strong discrepancies in all cases. These results indicate that the master equation description should hold for a much larger class of bath preparations in chaotic environments compared to integrable environments. Also the finite size scaling extracted in Fig.~\ref{fig:dynamics_sys_scaling} suggests that there are strong qualitative differences between chaotic and integrable baths.

In the integrable dynamics, we observe a weak form of equilibration (but not thermalisation) for all but the single eigenstate bath preparations. Our results therefore question whether thermalization in integrable models, as observed in Refs.~\cite{tasaki_macroscopic_2024_real,shiraishi_quantum_2024_real,PhysRevE.111.014129, usui_microscopic_2024-1, alba_quantum_2019_real, lai_entanglement_2015, cramer_quantum_2010-1, huerta_entropy_1969, hovhannisyan_long-time_2023_real, chakraborti_entropy_2022, baldovin_statistical_2021, cameo_2024}, are generic. Using the typicality arguments discussed in Sec.~\ref{sec:general_state}, we may expect that when the bath is large and prepared in a randomly sampled state it will appear thermal for local operators, independent of the details of the dynamics. Similar behaviour of the product state may be a result of sufficiently large participation of eigenstates. At finite size, the equilibrium state might be determined by the generalized Gibbs ensemble which is dependent on the energy density and the conserved charges of the model; however, this is left for future work.  Additionally, it is known that a smooth function in frequency can be extracted from the off-diagonals of matrix elements for operators in the basis of eigenstates of interacting integrable models at infinite temperature \cite{zhang_statistical_2022}, so it may be possible to extend the master equation derivation to these cases.

A variety of extensions to the results obtained above may be possible. First, using alternative assumptions to chaos for understanding thermalisation could be used such as typicality of pure states sampled from the Hilbert space~\cite{bartsch_occurrence_2008,PhysRevLett.102.110403, PhysRevE.97.062129}, discussed in Sec.~\ref{sec:general_state}, and typicality associated with the Hamiltonian~\cite{deutsch_quantum_1991, nation_off-diagonal_2018_real, Ithier_2017, PhysRevA.96.012108, bertoni_real}. It may also be possible to compute the spectral function for larger system sizes using typicality simulation methods~\cite{Heitmann_2020, PhysRevLett.128.180601}. In particular, the  correlation functions can be computed and Fourier transformed to obtain the spectral function, but this is left for future work. Our results could also be verified in experimental platforms with a high degree of tunability, where many-body integrability can be induced or removed. In Fig.~\ref{fig:populations}, we can observe qualitative differences in the behaviour of systems coupled to finite, integrable and non-integrable baths which are prepared in pure states. This could be experimentally tested by studying the behaviour of an impurity embedded in some isolated, many-body environment with tunable interactions. Experimental realizations of isolated chaotic and integrable many-body systems have been realized \cite{kinoshita_quantum_2006_real, gring_relaxation_2012_real, langen_local_2013_real, kaufman_quantum_2016, Rosenberg_2024, Keenan_2023, Scheie_2021, Jepsen_2020}, with ultracold Fermi gases representing a natural arena to study the behaviour of impurities which are interacting with environments that are chaotic and integrable. The ETH has been applied to transport scenarios with pure bath states~\cite{PhysRevA.107.022220, PhysRevA.105.L040203} and has been experimentally tested on superconducting quantum processors~\cite{zhang2024emergencesteadyquantumtransport}. It would be interesting to apply our results to transport phenomena of this kind.

\textit{Note added}. While this manuscript was being prepared, we became aware of the work of Purkayastha et al.~\cite{purkayastha2024differencethermalizationopenisolated} which also studies the relation between the ETH and OQS theory.

\begin{acknowledgments}
POD is supported by the Irish Research Council (ID GOIPG/2023/3847). PS is financially supported by ``la Caixa'' Foundation (ID 100010434, fellowship code LCF/BQ/PR21/11840014), the Ram\'on y Cajal program RYC2022-035908-I as well as the MICINN with funding from European Union NextGenerationEU (PRTR-C17.I1), by the Generalitat de Catalunya (project 2017-SGR-1127), the European Commission QuantERA grant ExTRaQT (Spanish MICIN project PCI2022-132965), and the Spanish MINECO (project PID2019-107609GB-I00) with the support of FEDER funds. KM is grateful for the support of the Australian Research Council Discovery Projects DP210100597. MTM is supported by a Royal Society-Science Foundation Ireland University Research Fellowship (URF\textbackslash R1\textbackslash 221571). 
JG is also funded by a Science Foundation Ireland-Royal Society University Research Fellowship.
We
acknowledge the Irish Centre for High End Computing and Kesha for the use of their computational facilities.
\end{acknowledgments}

\appendix
\renewcommand{\thesubsection}{\Roman{subsection}}

\section{Typical Microcanonical Pure States}
\label{Appendix:typicality}

In this section, we show that the vast majority of typical microcanonical pure states are locally equilibrated from the persepective of one- and two-point functions. This ensures that the stationarity condition in Eq.~\eqref{stationarity_cond} will be satisfied. We also derive an explicit expression for Fourier transform of the BCF for these states. typical microcanonical pure states are defined as a random superposition of eigenstates, $|n\rangle$, within a microcanonical energy window, $\mathcal{E}$, defined as the ensemble of energy eigenstates with energy $E_n \in [E_0 - \frac{\Delta E}{2}, E_0 + \frac{\Delta E}{2}]$.

Concentration of measures bound the probability that a function with a randomly sampled input will be far from the average of that function over all inputs. Levy's lemma states states that given a function $f: \mathbb{S}^{d} \rightarrow \mathbb{R}$, and a uniformly sampled element of a sphere $x \in \mathbb{S}^d$, the probability of $f(x)$ being outside some distance $\epsilon$ from the average over the hypersphere $\langle f\rangle$ is upper bounded as
\begin{equation}
\label{eq:levys_lemma}
    \text{Pr}(|f(x) - \langle f\rangle| > \epsilon) \leq 2\text{exp}\left({\frac{-2c(d+1)\epsilon^2}{\eta^2}}\right).
\end{equation}
Above, $c$ is a positive constant which can taken to be $(18\pi^3)^{-1}$, $d$ is the dimension of the hypersphere, and $\eta$ is the Lipschitz constant of the function $f$ which is a measure of the sensitivity of the function to its inputs and is given by $\eta = \text{sup}|\nabla f|$.

To use Levy's lemma, we first we compute the average of the one- and two-point functions over the microcanonical energy window. The microcanonical average of the one-point function is given by 
\begin{equation}
    \langle \hat{B}\rangle_{\text{MC}} = \frac{1}{d_{\mathcal{E}}} \sum_{n \in \mathcal{E}} \langle n|\tilde{B}(t)|n\rangle,
\end{equation}
with $d_{\mathcal{E}}$ the dimension of $\mathcal{E}$. Using the ETH, this can be written as
\begin{equation}
    \langle \hat{B}\rangle_{\text{MC}} = \frac{1}{d_{\mathcal{E}}} \sum_{n \in \mathcal{E}} \left[B(E_n) + e^{-\frac{S(E_n)}{2}} f(E_n,0) R_{nn}\right].
\end{equation}
Using the smoothness of the $f(E_n, 0)$ and $S(E_n)$ in energy, the extensiveness of $S(E_n)$ in system size, and the property that $\overline{R_{nm}} = 0$ the microcanonical average is given by 
\begin{equation}
    \langle \hat{B}\rangle_{\text{MC}} = \frac{1}{d_{\mathcal{E}}} \sum_{n \in \mathcal{E}} B(E_n).
\end{equation}
The smoothness of the diagonal term allows us to perform a Taylor expansion around the mean energy of the microcanonical window such that
\begin{equation}
    \langle \hat{B}\rangle_{\text{MC}} \approx B(E_0) + \frac{1}{2}\Delta E^2 B''(E_0),
\end{equation}
with $\Delta E$ the microcanonical energy fluctuations which we have assumed are subextensive. Note, we have used that the average energy of the microcanonical energy window is given by $E_0 = \frac{1}{d_{\mathcal{E}}}\sum_{n \in \mathcal{E}} E_n$. 

We perform a similar calculation for the two-point function by first noting that the BCF can be written in terms of the spectral function for a single eigenstate, as shown in Eq.~\eqref{BCF_ETH}. Using this result, the microcanonical average of the BCF is given by  
\begin{equation}
\begin{aligned}
    & \langle C(t, t')\rangle_{\text{MC}} = \frac{1}{d_{\mathcal{E}}}\sum_{n \in \mathcal{E}} \langle n|\tilde{B}(t) \tilde{B}(t')|n\rangle - \langle \tilde{B}(t)\rangle_{\text{MC}}^2\\ 
     &\qquad \qquad \quad \;  \approx \frac{1}{d_\mathcal{E}}\sum_{n\in \mathcal{E}}\int^{\infty}_{-\infty} d\omega e^{-i\omega \tau} e^{\beta \omega/2} |f(E_n, \omega)|^2.
\end{aligned}
\end{equation}
The temperature, $\beta = \beta(E_n)$, and the spectral function, $|f(E_n, \omega)|^2$, are both dependent on energy; however, they are insensitive to small changes in energy that occur within the energy window. We can Taylor expand these functions around $E_0$ and by neglecting the higher order terms we can write the microcanonical average of the BCF as
\begin{equation}
    \langle C(\tau)\rangle_{MC} \approx \int^{\infty}_{-\infty} d\omega e^{-i\omega \tau + \beta \omega/2} |f(E_0, \omega)|^2.
\end{equation}

We will now use Levy's lemma to bound the probability of a randomly sampled state looking non-thermal from the perspective of a one-point function. We use a known result that the Lipschitz constant for expectation values is upper bounded by the operator norm (or Schatten $\infty$-norm), denoted by $\|\cdot\|$, as \cite{popescu_foundations_2006_real} 
\begin{equation}
\label{lip_bound}
    \eta \leq 2 \|\hat{B}\|.
\end{equation}
Levy's lemma then gives us the following bound on the probability that the expectation value, $\langle \tilde{B}(t)\rangle_0 = \langle \psi_{B}|\tilde{B}(t)|\psi_{B}\rangle$, will be a distance $\epsilon$ away from the microcanonical average,
\begin{equation}
    \text{Pr}\left(|\langle \tilde{B}(t)\rangle_0 - \langle \tilde{B}(t)\rangle_{\text{MC}}| > \epsilon \right) \leq 2 \exp\left[-\frac{d_{\mathcal{E}}\epsilon^2}{18\pi^3\|\hat{B}\|^2}\right].
\end{equation}
Above, we have mapped the pure state of dimension $d_{\mathcal{E}}$ to a sphere of dimension $2d_{\mathcal{E}} - 1$ allowing us to use Eq \eqref{eq:levys_lemma}.

To use Levy's lemma on the BCF, we must find the Lipshitz constant for two-point correlation functions. We consider the real and imaginary parts separately:
\begin{equation}
\begin{aligned}
    &\text{Re}\left[C(\rho)\right] = \frac{1}{2}\text{tr}\left(\left\{\tilde{B}(t), \tilde{B}(0)\right\}\hat{\rho}\right),\\
    &\text{Im}\left[C(\rho)\right] = \frac{1}{2}\text{tr}\left(\left[\tilde{B}(t), \tilde{B}(0)\right]\hat{\rho}\right).
\end{aligned}
\end{equation}
The operator norm obeys the Cauchy-Schwarz inequality $\|\hat{A}\hat{B}\|\leq \|\hat{A}\|  \|\hat{B}\|$, and is invariant under unitary transformations $\|\hat{U}\hat{A}\hat{U}^{\dagger}\| = \|\hat{A}\|$. Using the result in Eq.~\eqref{lip_bound} and these properties, we can bound the Lipschitz constant for the total correlation function as
\begin{equation}
    \eta \leq 4 \|\hat{B}\|^2.
\end{equation}
From this we find an upper bound on the probability of the correlation function for a typical microcanonical pure state being outside a radius $\epsilon$ from the microcanonical average,
\begin{equation}
    \text{Pr}\left(|C(t, t') - \langle C(\tau)\rangle_{\rm MC}| > \epsilon\right) \leq  4\exp\left[-\frac{ d_{\mathcal{E}}\epsilon^2}{72\|\hat{B}\|^4}\right].
\end{equation}
The factor of $4$ arises due to the correlation function being complex.

We have computed a bound for the probability of a randomly sampled state from a microcanonical window being a distance $\epsilon$ from the microcanonical average for one- and two-point functions. We can see that provided $d_{\mathcal{E}}$ is sufficiently large and $\|\hat{B}\|$ is sufficiently small, then these functions will agree closely with the microcanonical average. The dimension of the microcanonical window will be exponentially large in system size and the local operator will be subextensive. The microcanonical averages are stationary so the conditions in Eq.~\eqref{stationarity_cond} will be satisfied. We have also found an exact form for the Fourier transform of the BCF in terms of the spectral function. This ensures that the analysis of Sec.~\ref{sec:energy_eigenstate} can be repeated to derive the Lindblad master equation.

\section{General Interaction Hamiltonian}
\label{Appendix:general_interaction}

In this section, we will show that the master equation can still be derived for eigenstates and typical microcanonical pure states when the interaction term between the system and bath is given by its general form
\begin{equation}
    \hat{V} = \sum_{\mu} \hat{S}^{\mu} \otimes \hat{B}^{\mu}.
\end{equation}
Above, $\hat{B}^{\mu}$ and $\hat{S}^{\mu}$ are the bath and system operator, respectively, and both are taken to be Hermitian operators without loss of generality. Given this interaction Hamiltonian, we will now have multiple BCFs given by
\begin{equation}
\label{eq:BCF_gen_int}
    C_{\mu \nu}(t, t') = \langle \tilde{B}^{\mu}(t) \tilde{B}^{\nu}(t')\rangle_0 - \langle \tilde{B}^{\mu}(t') \rangle_0 \langle \tilde{B}^{\nu}(t')\rangle_0.
\end{equation}

We will first consider an eigenstate bath preparation, $|n\rangle$, and then discuss the extension to typical microcanoncical pure states. Rewriting Eq.~\eqref{eq:BCF_gen_int} in terms of the energy eigenstates we get
\begin{equation}
\begin{aligned}
    &C_{\mu \nu}(t,t') = \sum_{m: m \neq n}e^{-i(E_m - E_n)t}e^{-i(E_n - E_m)t'} B^{\mu}_{nm} B^{\nu}_{mn}. 
\end{aligned}
\end{equation}
Inserting the ETH ansatz yields
\begin{equation}
\begin{aligned}
    &C_{\mu \nu}(\tau) = \sum_{m: m \neq n} e^{-i \omega_{nm}\tau} e^{-S(E_{nm})} R^{\mu}_{nm}R^{\nu}_{mn}\\
    & \qquad \qquad \qquad \times \left[f_{B}^{\mu}(E_{nm}, \omega_{nm}) f_{B}^{\nu}(E_{nm}, \omega_{nm})\right].
\end{aligned}
\end{equation}
Above, we have defined frequency $\omega_{nm} = E_m - E_n$, energy $E_{nm} = (E_n + E_m)/2$, and time $\tau = t - t'$.

Using identical assumptions from Sec.~\ref{sec:energy_eigenstate}: slow variance of $S(E)$ and $f^{\mu}(E, \omega)$ with energy, smoothness of the functions in the ETH ansatz, $t \delta E \ll 1$ we can fix an energy and frequency window within which we can replace the random matrix terms with their averages,
\begin{equation}
\begin{aligned}
    &C_{\mu \nu}(\tau) = \sum_{m: m \neq n} e^{-i \omega_{nm}\tau} e^{-S(E_{nm})} \overline{R^{\mu}_{nm}R^{\nu}_{mn}} \\
    & \qquad \qquad \qquad \times \left[f_{B}^{\mu}(E_{nm}, \omega_{nm}) f_{B}^{\nu}(E_{nm}, -\omega_{nm})\right].
\end{aligned}
\end{equation}
When $\mu = \nu$, we can use property that $\overline{|R^{\mu}_{nm}|^2} = 1$; however, when $\mu \neq \nu$ we must account for the potential correlations which may exist between different bath operators. 

Extensions to the ETH have shown that within narrow energy and frequency windows, the correlations between these different matrices can be characterized by a smooth function such that \cite{pappalardi_eigenstate_2022, noh_numerical_2020}
\begin{equation}
    \overline{R^{\mu}_{nm}R^{\nu}_{mn}} = h^{\mu \nu}(E_{nm}, \omega_{nm}).
\end{equation}
Using this ansatz, we can write the correlation function as
\begin{equation}
\label{eq:BCF_gen_int_discrete}
\begin{aligned}
    &C_{\mu \nu}(\tau) = \sum_{m: m \neq n} e^{-i \omega_{nm}\tau} e^{-S(E_{nm})} F^{\mu \nu}(E_{nm}, \omega_{nm})
\end{aligned}
\end{equation}
with the smooth function now defined as $F^{\mu \nu}(E, \omega) =  f_{B}^{\mu}(E, \omega) h^{\mu \nu}(E,  \omega) f_{B}^{\nu}(E, -\omega)$. 

As we have done in Sec.~\ref{sec:derivation}, we can replace the sum over energies in Eq.~\eqref{eq:BCF_gen_int_discrete} with an integral over the frequency,
\begin{equation}
    C_{\mu \nu}(\tau) = \int^{\max(\omega)}_{\min(\omega)} d\omega e^{-i\omega \tau} e^{\beta \omega /2} F^{\mu \nu}(E_n, \omega),
\end{equation}
Previously, we used that the spectral function decays on a frequency scale which is much smaller than $\max(\omega)$ and $\min(\omega)$. The function $h^{\mu \nu}(E, \omega)$ has been numerically studied in Ref.~\cite{noh_numerical_2020} and does not display the same decay properties as is seen in the ETH spectral function. Despite this, we postulate that function $F^{\mu \nu}(E, \omega)$ decays for positive and negative frequencies on a scale $\Delta^{+}_{\mu \nu} \ll \max(\omega)$ and $\Delta^-_{\mu \nu} \gg \min(\omega)$, respectively. This can be justified by the contributions of $f^{\mu}(E, \omega)$ and $f^{\nu}(E, -\omega)$ which decay exponentially, and assuming that $h^{\mu \nu}(E, \omega)$ does not grow exponentially with $\omega$. The correlation function can then be written as
\begin{equation}
    C_{\mu \nu}(\tau) = \int^{\infty}_{-\infty} d\omega e^{-i\omega \tau} e^{\beta \omega/2}F^{\mu \nu}(E_n, \omega).
\end{equation}

Using the same arguments from the simple interaction case, smoothness and decay of $F^{\mu \nu}(E, \omega)$ in frequency, we can justify the use of the Markov approximation, allowing us to write down the weak coupling Lindblad master equation: 
\begin{equation}
    \frac{d\hat{\rho}_S}{dt} = -i\left[\hat{H}_S', \hat{\rho}_S\right] + \sum_{\mu, \nu, \omega}\gamma_{\mu \nu}(E, \omega)\mathcal{D}[\hat{S}^{\mu}(\omega), \hat{S}^{\nu}(\omega)]\hat{\rho}_S.
\end{equation}
Here, the dissipator is given by $\mathcal{D}[\hat{A}, \hat{B}] \bullet= \hat{A}\bullet \hat{B}^{\dagger} - \frac{1}{2}\{\hat{B}^{\dagger}\hat{A}, \bullet\}$ and the transition rate matrix is given by $\gamma_{\mu\nu}(E, \omega) = \kappa^2 \int^{\infty}_{-\infty} d\tau e^{-i\omega \tau}C_{\mu \nu}(\tau) = 2\pi \kappa^2 e^{\beta \omega/2}F^{\mu \nu}(E, \omega)$.

For the above equation to produce completely positive, trace-preserving dynamics, we require $\gamma_{\mu\nu}(E, \omega)$ to be a positive, semidefinite matrix---have positive eigenvalues. To show this result, we first use the property that when the operators $\hat{S}^{\mu}$ are Hermitian, then the spectral function is Hermitian,
\begin{equation}
\label{gen_int_spec_func_properties}
    F^{\mu \nu}(E, \omega) = [F^{\nu \mu}(E, \omega)]^{*}.
\end{equation} 
This property ensures that there exists a unitary operator, $U^{\bar{\mu} \mu} \equiv U^{\bar{\mu} \mu}(E, \omega)$, which diagonalizes the transition rate matrix,
\begin{equation}
\begin{aligned}
    &\tilde{\gamma}_{\bar{\mu} \bar{\nu}} (E, \omega) \delta_{\bar{\mu}\bar{\nu}} = [U^{\dagger}]^{\bar{\mu} \mu}\gamma_{\mu \nu}(E, \omega)[U^{\nu \bar{\nu}}]\\
    & = \kappa^2 \int^{\infty}_{-\infty} d\tau e^{-i\omega \tau} [U^{\dagger}]^{\bar{\mu} \mu} C_{\mu \nu}(\tau) U^{\nu \bar{\nu}}.
\end{aligned}
\end{equation}
We can now define the diagonalized correlation matrix with respect to the transformed set of bath operators $\tilde{A}_{\bar{\mu}}(\tau) = \sum_{\mu} [U^{\dagger}]^{\bar{\mu} \mu} \tilde{B}_{\mu}(\tau)$ and $[\tilde{A}_{\bar{\nu}}(0)]^{\dagger} = \sum_{\nu} \tilde{B}_{\nu}U^{\nu\bar{\nu}}$
\begin{equation}
    D_{\bar{\mu} \bar{\nu}}(\tau) \delta_{\bar{\mu} \bar{\nu}} = \langle \tilde{A}_{\bar{\mu}}(\tau) \tilde{A}_{\bar{\nu}}(0)^{\dagger}\rangle_0 - \langle \tilde{A}_{\bar{\mu}}(\tau) \rangle_0 \langle \tilde{A}_{\bar{\nu}}(0)^{\dagger} \rangle_0.
\end{equation}
The operators $\tilde{A}_{\bar{\mu}}$ are a sum of local operators and therefore are expected to obey the ETH. Applying the ETH ansatz to these operators, we can write the diagonalized correlation matrix as
\begin{equation}
    D_{\bar{\mu} \bar{\mu}}(\tau) = \int^{\infty}_{-\infty} d\omega e^{-i\omega \tau} e^{\beta \omega /2} |f_{\bar{\mu}}(E, \omega)|^2.
\end{equation}
The eigenvalues of the transition matrix are given by
\begin{equation}
    \tilde{\gamma}_{\bar{\mu} \bar{\mu}}(E_n, \omega) = 2\pi \kappa^2 e^{\beta\omega/2} |f_{\bar{\mu}}(E_n, \omega)|^2. 
\end{equation}
These are non-negative, ensuring that the dynamics is described by a completely positive and trace-preserving map, and obey local detailed balance condition due to the symmetry property $|f_{\bar{\mu}}(E, \omega)|^2 = |f_{\bar{\mu}}(E, -\omega)|^2$. The Lindblad master equation can be rewritten as
\begin{equation}
    \frac{d \hat{\rho}_S}{dt} = -i[\hat{H}'_S, \hat{\rho}_S] + \sum_{\bar{\mu}, \omega}\tilde{\gamma}_{\bar{\mu} \bar{\mu}}(E, \omega) \mathcal{D}[\hat{K}_{\bar{\mu}}(\omega)]\hat{\rho}_{S},
\end{equation}
where we have redefined the system operators as $\hat{K}_{\bar{\mu}}(\omega) = \sum_{\mu} [U^{\dagger}(E,\omega)]^{\bar{\mu} \mu} \hat{S}_{\mu}(\omega)$ which can still be interpreted as raising and lowering operators $[\hat{H}'_S, \hat{K}_{\bar{\mu}}(\omega)] = -\omega \hat{K}_{\bar{\mu}}(\omega)$.

So far, we have shown that a bath prepared in a single eigenstate and for a general interaction Hamiltonians we can derive an explicit form for the Fourier transform of the BCF which ensures the validity of the master equation. To extend this analysis to typical microcanonical pure states, we note that we can simply apply the analysis of App. \ref{Appendix:typicality} to each of the correlation functions $D_{\bar{\mu} \bar{\mu}}(\tau)$. It is straightforwardly seen that the master equation will hold for these states.

\section{Thermodynamics quantities from the density of states}
\label{temperature_Appendix}

\begin{figure}[h]
    \centering
    \includegraphics[width=0.47\textwidth]{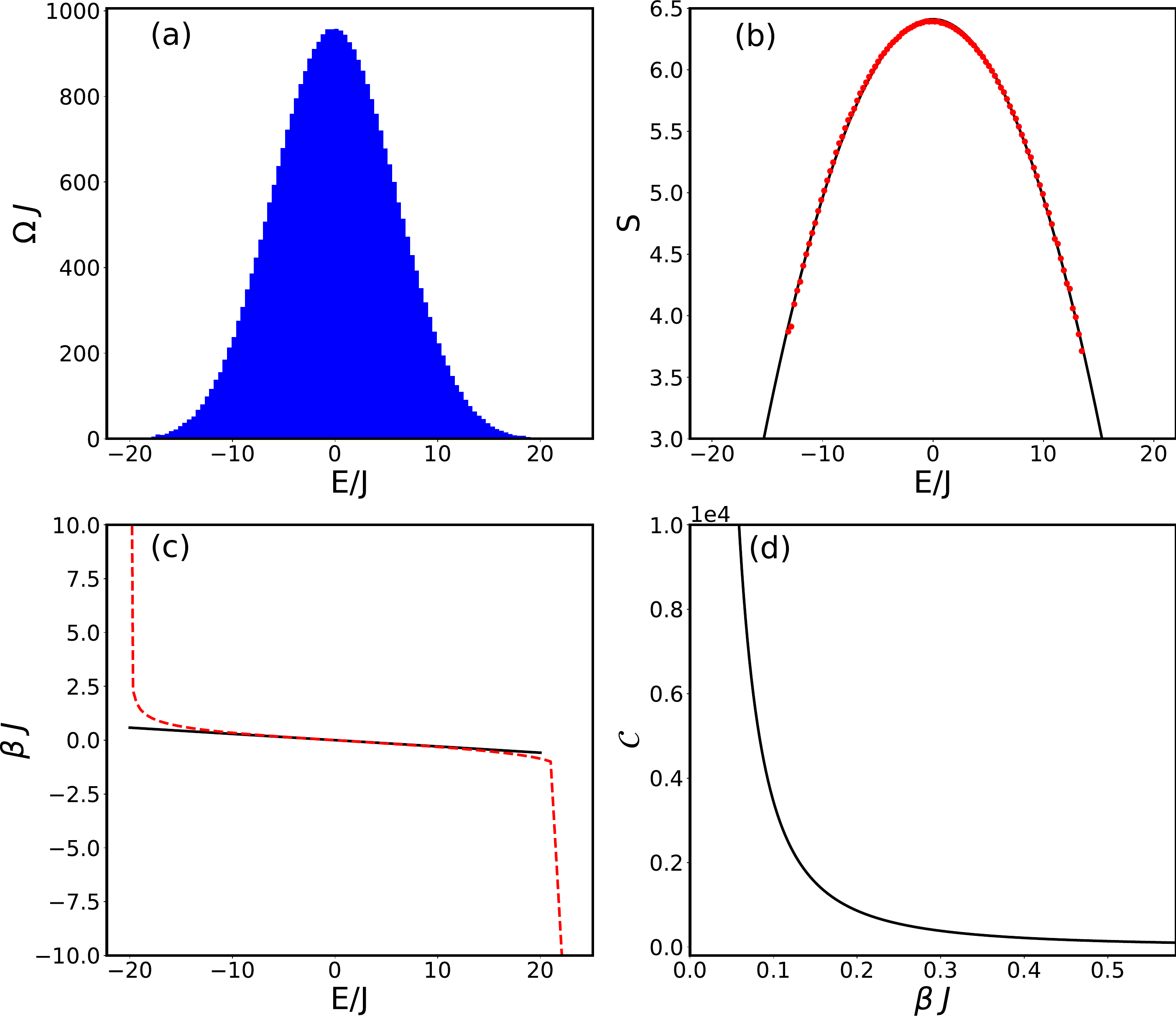}
    \caption{(a) The density of states for the chaotic Ising model with $L = 15$ sites. (b) The Boltzmann entropy computed by taking the logarithm of the density of states. We perform a fit of this data to a quadratic polynomial. [(c), (d)] These are the temperature and heat capacity computed using the quadratic fit of the Boltzmann entropy.  We also compare the temperature computed using the canonical ensemble and see close agreement near the center of the spectrum due to ensemble equivalence.}
    \label{fig:dos_c}
\end{figure}

\begin{figure}[t]
    \centering
    \includegraphics[width=0.47\textwidth]{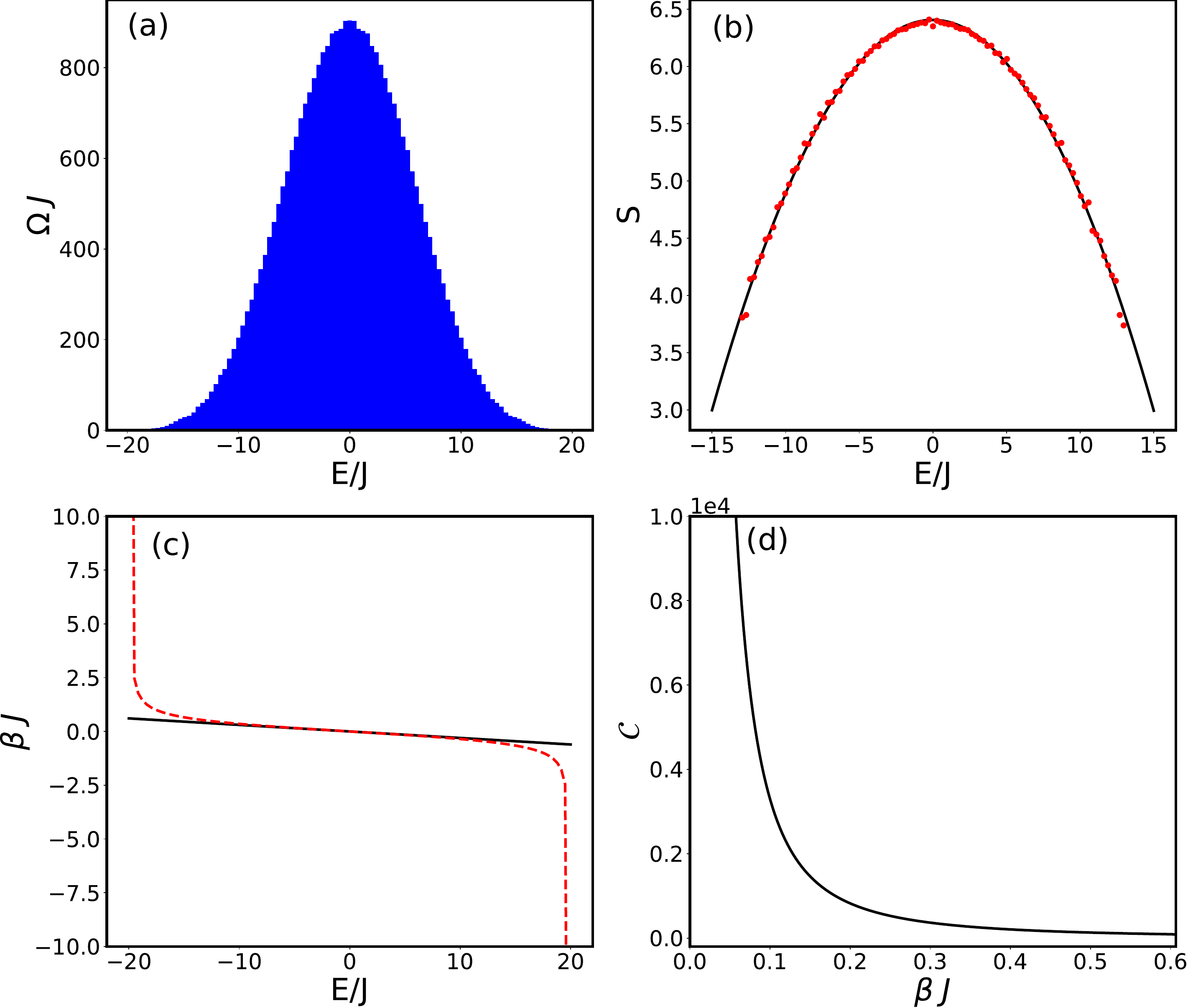}
    \caption{(a) The density of states for the integrable Ising model with $L = 15$ sites. (b) The Boltzmann entropy computed by taking the logarithm of the density of states. We perform a fit of this data to a quadratic polynomial. [(c), (d)] These are the temperature and heat capacity computed using the quadratic fit of the Boltzmann entropy.  We also compare the temperature computed using the canonical ensemble and see close agreement near the center of the spectrum due to ensemble equivalence.}
    \label{fig:dos_i}
\end{figure}

Throughout this paper we have assigned a variety of thermodynamic quantities, such as Boltzmann entropy, temperature, and heat capacity, to pure states. In this section we will discuss how we compute these quantities. We have exactly diagonalized the bath Hamiltonian in this paper and so we have access to full spectrum. We can divide the spectrum into narrow energy windows of width $\Delta E$ which is larger then the level spacing statistics, but narrower then the total bandwidth of the spectrum. Counting the number of states within each window gives us the density of states, $\Omega(E)\Delta E$, from which all thermodynamics quantities can be computed. In Figs. \ref{fig:dos_c}a and \ref{fig:dos_i}a, we plot the density of states for the chaotic and integrable models and observe a Gaussian distribution which is expected for locally interacting many-body Hamiltonians. 

The Boltzmann entropy is computed from the density of states as
\begin{equation}
    S(E) = \text{log}(\Omega(E) \Delta E).
\end{equation}
From standard thermodynamics, the inverse temperature and heat capacity are given by Eq.~\eqref{eq:temp_heat}.
We compute the temperature and heat capacity by performing a polynomial fit of the logarithm of the density of states and compute the derivative of the outputs, shown in Figs. \ref{fig:dos_c} and \ref{fig:dos_i}. We also compute the canonical temperature of a pure state $|\psi\rangle$ by solving the following equation for $\beta$
\begin{equation}
    \langle \psi| \hat{H}|\psi\rangle = \text{tr}\left(\hat{H}\frac{e^{\beta \hat{H}}}{\mathcal{Z}}\right).
\end{equation}
Comparing the microcanonical and the canonical temperatures, we see that the two approach's match reasonably well up until the edge of the spectrum which is a result of ensemble equivalence. We note that a variety of other approaches can be taken to compute the temperature for pure states \cite{burke_entropy_2023}.

\section{Level Spacing Statistics of the Total Hamiltonian}
\label{level_statistics}

\begin{figure}[b]
\centering
\includegraphics[width=0.48\textwidth]{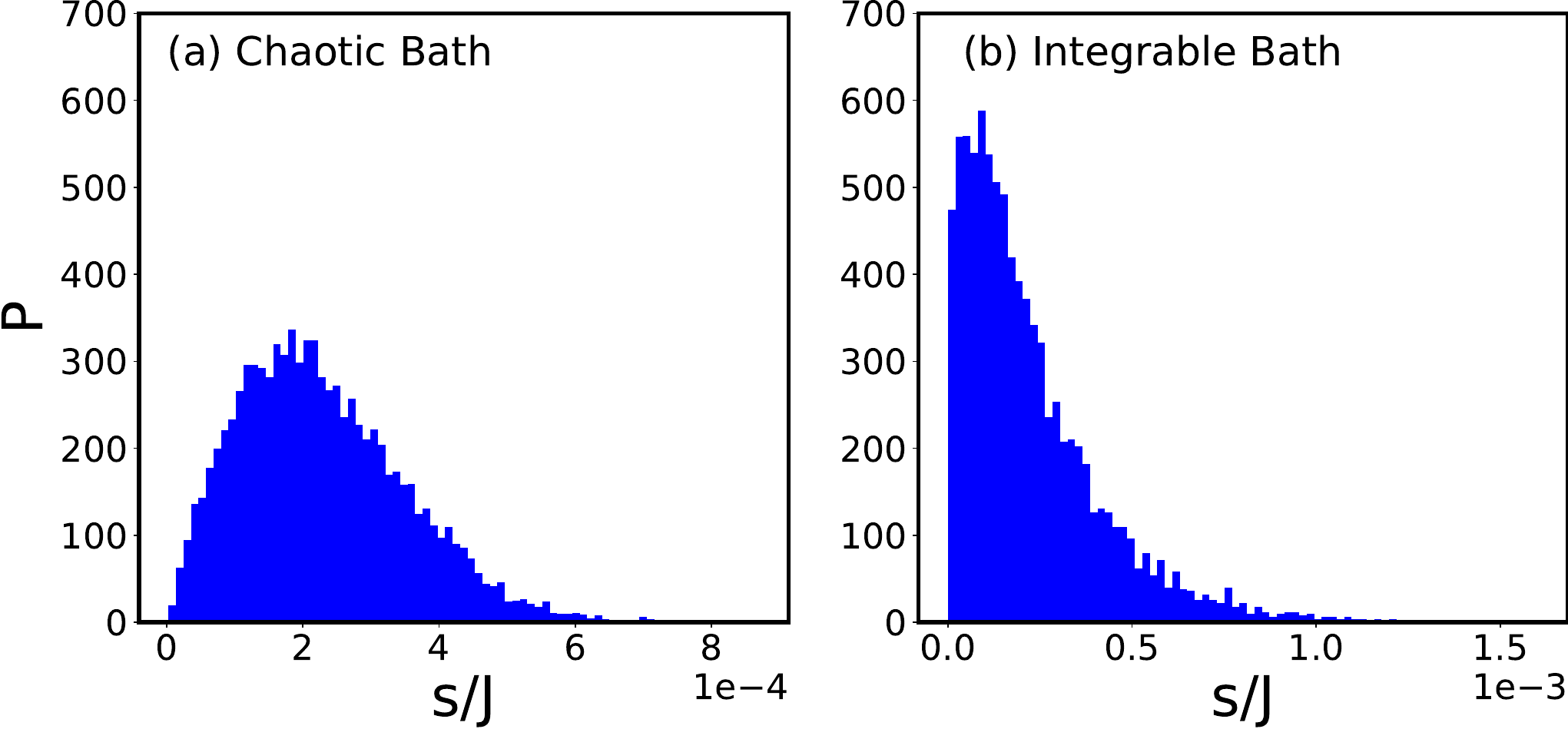}
\caption{The level spacing statistics for the eigenvalues of the entire system and bath Hamiltonian. (a) For a chaotic bath, we see level repulsion and a distribution which agrees with Wigner-Dyson statistics indicating chaotic behaviour. (b) For the integrable bath, we observe level spacing statistics that is inconsistent with the Wigner-Dyson distribution suggesting that the total Hamiltonian interaction term has not broken integrability.}
\label{fig:level_stats}
\end{figure}

In this section we discuss the behaviour of the total system and bath Hamiltonian, $\hat{H} = \hat{H}_{S} + \hat{H}_{B} + \kappa \hat{V}$. It is known that interacting integrable systems are extremely sensitive to perturbations such that even a single impurity on a site can break the conserved charges of the system \cite{santos_speck_2020, brenes_eigenstate_2020, brenes_low-frequency_2020}. Given this sensitivity, we will now check whether the total Hamiltonian is integrable when introducing the coupling between the system and bath.

We use level spacing statistics to diagnose whether the total Hamiltonian is chaotic \cite{dalessio_quantum_2016}. This is a well established probe for chaotic behaviour in quantum systems and acts as a probe of the number of conserved quantities in a model. Chaotic systems should have no degeneracies as these models will not have any conserved charges, other then energy conservation. The level statistics of chaotic systems will be described by a Wigner-Dyson distribution which displays level repulsion. Non-chaotic models will have degeneracies and so its level statistics will not be described by the Wigner-Dyson distribution.

In Fig. \ref{fig:level_stats}, we see the level spacing statistics of the full system and environment Hamiltonian for the chaotic and integrable baths. In the chaotic case we see standard Wigner-Dyson statistics with level repulsion due to the lack of conserved charges. In the integrable bath case, we observe that the interaction term does not break integrability as the level spacing distribution displays degeneracies. However, we also note that integrable systems are expected to have exponential level spacing distribution which is not shown here. The bath Hamiltonian is a transverse field Ising model which is a non-interacting so we expect these models to be more robust against perturbations.

\bibliography{main_draft_v2}

\end{document}